\documentclass{aa}  
 \usepackage{natbib}                                   
 \usepackage{upgreek}
\usepackage{graphicx}
\usepackage{amssymb}
\usepackage{txfonts}
\usepackage{color}
\usepackage{enumerate}
\usepackage{gensymb}  
\usepackage{natbib,twoopt}
\usepackage[breaklinks=true]{hyperref} 
\usepackage{cancel,soul,ulem}  

\bibpunct{(}{)}{;}{a}{}{,} 
\makeatletter
\newcommandtwoopt{\citeads}[3][][]{\href{http://adsabs.harvard.edu/abs/#3}%
{\def\hyper@linkstart##1##2{}%
\let\hyper@linkend\@empty\citealp[#1][#2]{#3}}}
\newcommandtwoopt{\citepads}[3][][]{\href{http://adsabs.harvard.edu/abs/#3}%
{\def\hyper@linkstart##1##2{}%
\let\hyper@linkend\@empty\citep[#1][#2]{#3}}}
\newcommandtwoopt{\citetads}[3][][]{\href{http://adsabs.harvard.edu/abs/#3}%
{\def\hyper@linkstart##1##2{}%
\let\hyper@linkend\@empty\citet[#1][#2]{#3}}}
\newcommandtwoopt{\citeyearads}[3][][]%
{\href{http://adsabs.harvard.edu/abs/#3}
{\def\hyper@linkstart##1##2{}%
\let\hyper@linkend\@empty\citeyear[#1][#2]{#3}}}
\makeatother
\def\hi{H{\textsc i} }
\def\h2{H{\textsc 2} }
\def\co{$^{12}$CO }
\def\co13{$^{13}$CO }
\def\co18{C$^{18}$O }
\def\cmthr{cm$^{-3} $ }
\def\cmtwo{cm$^{-2}$ }

\def\r{\mathrm}
\def\i{\textit}

\def\arcsec{\hbox{$^{\prime\prime}$}}
\def\arcmin{\hbox{$^{\prime}$}}

\def\cm2{cm$^{-2}  $}

\def\kms{km~s$^{-1}$}
\def\nh3{NH$_3$}
\def\n2h{N$_2$H$^+$}
\def\hc3n{HC$_3$N}
\def\h2{H$_2$}
\def\nh{n(H$_2$)}
\def\cp{C$^+$  }

\begin{document}

\title{Physical properties of CO-dark molecular gas traced by C$^+$}

\author{Ningyu Tang\inst{1,2}, Di Li \inst{1,3}, Carl Heiles \inst{4}, Shen Wang \inst{1,2}, Zhichen Pan 
\inst{1,2},  Jun-Jie Wang\inst{1}}

\authorrunning{N.Y. Tang et al.}
\titlerunning{Physical properties of CO-dark molecular gas traced by C$^+$}

\institute{National Astronomical Observatories, CAS, Beijing 100012, China; \email{nytang@nao.cas.cn, dili@nao.cas.cn}
\and  
University of Chinese Academy of Sciences, Beijing 100049, China 
\and 
Key Laboratory of Radio Astronomy, Chinese Academy of Sciences 
\and
Astronomy Department, University of California, Berkeley }

\abstract
{Neither H{\sc i} nor CO emission can reveal a significant quantity of so-called  dark gas in the interstellar medium (ISM). It is considered that CO-dark molecular gas (DMG), the molecular gas with no or weak CO emission, dominates dark gas.  Determination of physical properties of DMG is critical for understanding ISM evolution.  Previous studies of  DMG in the Galactic plane  are based on assumptions of excitation temperature and volume density. Independent measurements of temperature and volume density are necessary.}
{We intend to  characterize physical  properties of DMG  in the Galactic plane based on \cp  data from the $Herschel$ Open Time Key Program, namely Galactic Observations of Terahertz C+ (GOT C+) and \hi narrow self-absorption (HINSA) data from international \hi 21 cm Galactic plane surveys.}
{We identified DMG clouds with HINSA features  by comparing \hi, C$^+$, and CO spectra. We derived the \hi excitation temperature and \hi column density through spectral analysis of HINSA features. The \hi volume density was determined by utilizing the on-the-sky dimension of the cold foreground \hi cloud under the assumption of axial symmetry.  The column and volume density of \h2 were derived through excitation analysis of \cp emission. The derived parameters were then compared with a chemical evolutionary model.}
 {We identified 36 DMG clouds with HINSA features. Based on uncertainty analysis, optical depth of \hi  $\tau\rm_{\hi}$ of 1 is a reasonable value for most clouds. With the assumption of $\tau\rm_{\hi}=1$, these clouds were  characterized by excitation temperatures in  a range of 20 K to 92 K with a median value of 55 K and volume densities in the range of $6.2\times 10^1$ \cmthr to $1.2\times10^3$ \cmthr with a median value of $2.3\times 10^2$ \cmthr.   The fraction of DMG column density in the cloud ($f\rm_{DMG}$) decreases with increasing excitation temperature following an empirical relation $f\r{_{DMG}}=-2.1\times 10^{-3}T\r{_{ex},(\tau\rm_{\hi}=1)}+1.0$.  The relation between  $f\rm_{DMG}$ and  total hydrogen column density $N\rm_H$ is given by $\rm \i{f}_{DMG}=1.0-3.7\times 10^{20}/\i{N}_H$. We divided the clouds into a high extinction group and low extinction group with the dividing threshold being total hydrogen column density $N\rm_H$ of $5.0\times 10^{21}$  \cmtwo  ($A\rm_V = 2.7$ mag).  The values of $f\rm_{DMG}$ in the low extinction group ($A\rm_V \le 2.7$ mag)  are consistent with the results of the time-dependent, chemical evolutionary model at the age of $\sim$10 Myr. Our empirical relation cannot be explained by the chemical evolutionary model for clouds in the high extinction group ($A\rm_V > 2.7$ mag). Compared to clouds in the low extinction group ($A\rm_V \le 2.7$ mag), clouds in the high extinction group ($A\rm_V > 2.7$ mag)  have comparable volume densities but excitation temperatures that are 1.5 times lower. Moreover, CO abundances in clouds of the high extinction group ($A\rm_V > 2.7$ mag) are $6.6\times 10^2$ times smaller than the canonical value in the Milky Way.}
 { The molecular gas seems to be the dominate component in these  clouds. The high percentage of DMG in clouds of the high extinction group ($A\rm_V > 2.7$ mag) may support the idea that molecular clouds are forming from pre-existing molecular gas, i.e., a cold gas with a high H$_2$ content but that contains a little or no CO content.}

 
   \maketitle
   
\section{Introduction}
\label{sec:introduction}

The interstellar medium (ISM) is one of the fundamental baryon components of galaxies. The ISM hosts star formation. Determining the composition of the ISM will improve understanding of the lifecycle of ISM and the evolution of galaxies.

The 21-cm hyperfine line of atomic hydrogen has been used to trace the neutral medium. The linear relation between H{\sc i} column density and visual extinction, $N(\mathrm {\hi})/A\rm_V=1.9\times 10^{21}\:cm^{-2} mag^{-1}$ (Bohlin, Savage, \& Drake 1978), is valid for $A\rm_V < 4.7$ mag. Molecular hydrogen, \h2, the main component of ISM,  lacks a permanent dipole moment and does not have   rotational radio transitions  in cold ISM. CO and its isotopologues have been used as the main tracers of dense, well-shielded \h2 gas. At  Galactic scales, \h2 column density is derived through multiplying  integrated CO intensity $W(\rm CO)$ by an $X_{\rm CO}$ factor of $\rm 2\times 10^{20} cm^{-2}(K \cdot km s^{-1})^{-1}$ with $\pm 30\%$ uncertainty in the Milky Way disk (Bolatto, Wolfire, \& Leroy 2013). Though mean values of $X_{\rm CO}$ are similar for CO-detected molecular gas in diffuse and dense environments (Liszt et al. 2010), the volume density of CO-detected molecular gas is an order of magnitude greater than typical values of diffuse atomic gas shown in Heiles \& Troland (2003). 

The transition from diffuse atomic hydrogen to dense CO molecular gas  is not well understood. Dust is assumed to be mixed well with gas. Infrared emission  of dust has  been used as a tracer of total hydrogen column density. Results from the all-sky infrared survey by the Infrared Astronomical Satellite (IRAS), the Cosmic Background Explorer (COBE), and the Planck satellite revealed excess of dust emission, implying additional gas that cannot be accounted by H{\sc i} and CO alone (Reach, Koo, \& Heiles 1994; Reach, Wall, \& Odegard 1998; Hauser et al.\ 1998; Planck Collaboration 2011). Furthermore, the gamma-ray observations from COS-B (Bloemen et al.\ 1986) and the Energetic Gamma-Ray Experiment Telescope (EGRET; Strong \& Mattox 1996; Grenier, Casandjian, \& Terrier 2005)  also implied an extra gas component with a mass comparable to that in gas traced by $N(\rm \hi)$+2$X\rm _{CO}$*$W(\rm CO)$ in the Milky Way. This excess component of ISM, which cannot be fully traced by the usual H{\sc i} 21-cm  or CO 2.6-mm transition, is termed  dark gas. 

 The mainstream view considers dark gas to be unobserved molecular gas due to lack of corresponding CO emission.  Most of the direct detections of molecular gas were made with CO emission. There are, however, examples of interstellar molecules detected toward lines of sight without corresponding CO emissions (Wannier et al.\ 1993; Magnani \& Onello 1995;  Allen et al.\ 2015). If the nondetection of CO is taken as a sign of missing molecular gas, the fraction of dark gas varies from 12\% to 100\% for individual components in Liszt \& Pety (2012). The existence of unresolved molecular gas by CO  is also supported by photodissociation region (PDR) model 
 (e.g., van Dishoeck et al.\ 1988). H$_2$ can exist  outside the CO region in an illuminated cloud because the self-shielding threshold of H$_2$ is smaller than that of CO.  The gas in the transition layer between the outer H$_2$ region and the CO region is  CO-dark molecular gas (DMG). 

The DMG can be  associated with H{\sc i} self-absorption (HISA), which is caused by a foreground H{\sc i} cloud that is colder than H{\sc i} background at the same radial velocity (e.g., Knapp et al.\ 1974).  The Canadian Galactic Plane Survey (CGPS; Gibson et al.\ 2000, Taylor et al.\ 2003) and the Southern Galactic Plane Survey (SGPS; McClure-Griffiths et al.\ 2005) revealed that HISA is correlated with molecular emission in space and velocity (Gibson et al.\ 2005b, Kavars et al.\ 2005) although  HINSA without \h2 can exist (Knee \& Brunt 2001). It is now accepted fully that  large portion of cold neutral medium (CNM) is colder (Heiles \& Troland 2003) than the predictions of the three-phase ISM model (McKee \& Ostriker 1977). When  HINSA does contain \h2, it is dubbed \hi narrow self-absorption (HINSA; Li \& Goldsmith 2003). In normal molecular clouds, HINSA can be easily identified through its correlation with $^{13}$CO.  Without the emission of CO as a clear comparison mark, distinguishing between HISA and HINSA relies on the empirical threshold of $\delta V \sim$ 1.5 km/s, which seems to be applicable in most diffuse regions, but can be subjective. We  henceforth adopt the term HINSA  because of our focus on DMG. 

 The total H$_2$ column density  can be measured directly through ultraviolet (UV) absorption of H$_2$ toward stars. Observations  taken by the $Copernicus$ satellite (Savage et al.\ 1977) and the \i{Far Ultraviolet Spectroscopic Explorer (FUSE)} satellite (Rachford et al.\ 2002, 2009) revealed a  weak inverse correlation between rotational temperature and reddening, as well as  increasing correlation between molecular fraction and reddening. These kinds of observations are limited to strong UV background stars with low extinction ($<$ 3 mag), and cannot resolve a Galactic cloud due to coarse spectral resolution ($>$ 10 \kms) at UV bands (Snow \& McCall 2006).  The C$^+$-158  $\mu$m emission, a fine structure transition $^2P_{3/2}$$\rightarrow$$^2P_{1/2}$, can be used as a probe of molecular gas in PDR. Based on \cp spectra obtained  from a $Herschel$ Open Time Key Program, Galactic Observations of Terahertz C+ (GOT C+),  Langer et al.\ (2014; hereafter  L14)  found that DMG mass fraction varies from $\sim 75\%$ in diffuse molecular clouds without corresponding CO emission  to $\sim 20\%$ for dense molecular clouds with CO emission. 

There are two critical challenges in quantifying the DMG environment: the determination of the kinetic temperature, $T\rm_{k}$, and the determination of  the column and volume densities of \hi and \h2. Analysis of dust emission and extinction can aid in meeting these challenges. When looking into the Galactic plane, however, analysis of dust is muddied by source confusion. In previous studies, Kavars et al.\ (2005) attempted to constrain $T\rm_{k}$ and volume density $n$ based on an analysis of \hi absorption. Because of the lack of an effective tracer of total hydrogen gas, these authors had to rely on the Galactic thermal pressure distribution in Wolfire (2003) to estimate the molecular fraction. The study L14 introduced \cp emission as an effective tracer of total hydrogen gas. The study L14 assumed an overall  temperature of 70 K and Galactic thermal pressure distribution to calculate total volume density. They analyzed \cp excitation to determine molecular abundance. Because \cp emission is sensitive to kinetic temperature and volume density, the  lack of direct measurements of excitation temperature and volume density in L14  introduced uncertainties, especially for single clouds at low Galactic latitudes. Moreover,  L14 obtained H{\sc i} intensity by integrating velocity width  defined by the \cp (or $^{13}$CO) line. This overestimated  H{\sc i} column density as widespread background \hi emission is  included. Additionally, the optically thin assumption for the 21-cm line adopted in L14 results in an uncertainty of 20\%  for optical depth between 0.5 and 1 as discussed in their paper. Considering the  caveats above,  it is of great importance to inspect the effects of kinetic temperature, volume density, and \hi optical depth.
 
To improve  constraints on physical properties of  DMG, we adopted here the HINSA method in Li \& Goldsmith (2003) to obtain an independent measure of $T\rm_{ex}(\hi)$, $N(\rm\hi)$, and $\rm \i{n}(\hi)=\i{N}(\hi)/\i{L}_{\hi}$, where $L\rm_{\hi}$ is the linear dimension of HINSA cloud.  \h2 volume density, $n$(\h2), and \h2 column density, $N$(\h2), can be related as $\rm \i{n}(H_2)=\i{N}(H_2)/\i{L}\rm_{H_2}$, where $L\rm_{H_2}$ is the linear dimension of H$_2$ region of the HINSA cloud. According to the PDR model T6 (model parameters are proton density of 10$^3$ cm$^{-3}$, temperature of 15 K, UV intensity of 1, and total visual extinction of 5.1 mag) in van Dishoeck \& Black (1988), the outer layer with pure \hi ($L\rm_{\hi}$$-L\rm_{H_2}$ $\sim$ 0.03 pc) is relatively thin compared to the cloud with A$\rm_V=1$ mag ($\sim$ 0.9 pc). $N$(\h2) and $n$(\h2) can then be determined through \cp excitation analysis after adopting the ratio $L\r{_{H_2}}/L\rm_{\hi}=1$. The uncertainty caused by the value of $L\r{_{H_2}}/L\rm_{\hi}$ is discussed in detail in Section \ref{sec:cpanalysis}.

This paper is organized as follows. In Section \ref{sec:observations}, we describe our observations and data. In Section \ref{sec:analysis}, we present our procedure to identify DMG clouds. In Section \ref{sec:results}, we present derived  spatial distribution, \hi excitation temperatures, column  and volume densities of \hi and \h2 of  identified DMG clouds   from \hi and \cp analysis. In Section \ref{sec:DMGcloud},  we present derived DMG cloud properties.  The discussion and summary are presented in Section \ref{sec:discussion} and Section \ref{sec:summary}, respectively.


\section{Data }
\label{sec:observations}

\subsection{$\rm{C^+}$}
\label{sec:cplus}
The $Herschel$ Open Time Key Program, GOT C+  observed C$^+$-158 $\mu$m line toward 452 lines of sight toward the Galactic plane (e.g., Langer et al.\ 2010). Most of the lines of sight  are within 1 degree of the Galactic plane in latitude except for a small fraction of lines of sight in the outer Galaxy that are within 2 degrees. Longitude distribution of all lines of sight  can be  found in Figure 1 of L14. 
We obtained public \cp data from $Herschel$ Science Archive (HSA) with the kind aid of J. Pineda. The angular resolution of the \cp observations is 12\arcsec. The data have already been smoothed to a channel width of 0.8 km s$^{-1}$ with an average root mean square (rms) of 0.1 K. A detailed description of the GOT C+ program and the data can be found in Pineda et al.\ (2013) and  L14.

\subsection{\rm \hi}
\label{sec:HI}

The H{\sc i} data used here were taken from international H{\sc i} Galactic plane surveys. The Southern Galactic Plane Survey (SGPS; McClure-Griffiths et al.\ 2005) covers Galactic longitudes 253\degree$\le l \le$358\degree\ and  5\degree $\le l \le$20\degree\ with latitudes $|b|\le 1.5 \degree$.\ The SGPS data have an angular resolution of 2\arcmin\ and a rms of 1.6 K per 0.8 km s$^{-1}$. The VLA Galactic Plane Survey (VGPS; Stil et al.\ 2006) covers Galactic longitudes 18\degree $\le l \le$67\degree\ and latitudes $|b|\le 1.3 \degree$\ to $2.3 \degree$.\  The VGPS image data have an angular resolution of 1\arcmin \   and a rms  of 2 K per 0.82 km s$^{-1}$.  The Canadian Galactic Plane Survey (CGPS; Taylor et al.\ 2003) covers Galactic longitudes 63\degree $\le l \le$175\degree\ and latitude  -3.6\degree $\le b \le$5.6\degree.\  The CGPS data have an angular resolution of 1\arcmin\ and a rms of 2 K per 0.82 km s$^{-1}$. The Galactic Arecibo L-band Feed Array H{\sc i}  (GALFA-H{\sc i}; Peek et al.\ 2011) survey covers  -1.3\degree $\le \delta \le$38.0\degree,  \ about  32\% of the sky. For Galactic longitudes 180\degree $\le l \le$ 212\degree,\  we extracted  H{\sc i} spectra from GALFA-H{\sc i} data with an angular resolution of 3.4\arcmin\ and a rms of 80 mK per 1 \kms.

\subsection{$\rm{CO} $}
\label{sec:CO}
For lines of sight of GOT C+ in the Galactic longitude -175.5\degree$\le l \le$  56.8\degree\ ,  $J=1$ $\rightarrow$ 0 transitions of $^{12}$CO, $^{13}$CO, and C$^{18}$O were observed  with  the ATNF Mopra Telescope  (see Pineda et al.\  2013 and L14 for details). The Mopra data have an angular resolution of 33\arcsec. Two channels of CO spectrum were smoothed into one to derive a comparable velocity resolution to that of  \hi spectra.  The typical rms values are  0.44 K for $^{12}$CO per  0.7 km s$^{-1}$,  0.18 K for $^{13}$CO per  0.74  km s$^{-1}$,  and 0.21 K for C$^{18}$O per 0.73 km s$^{-1}$.

For those GOT C+ sightlines (56.8$\degree$ $<l<$  184.5$\degree$) that are out of the Mopra sky coverage, we obtained $J=1\rightarrow 0$ transitions of $^{12}$CO, $^{13}$CO, C$^{18}$O with the Delingha 13.7 m telescope. Full width at half power of Delingha telescope is about 60\arcsec. The observations were made between May 9 and 14 2014, using the configuration of 1 GHz bandwidth and 61 kHz channel resolution (velocity resolution of $\sim 0.16$ km s$^{-1}$).  The data were reduced with {\tt GILDAS/CLASS}\footnote{{\tt
http://www.iram.fr/IRAMFR/GILDAS}} data analysis  software and were smoothed to $\sim$0.8 km s$^{-1}$ to be consistent with velocity resolution of H{\sc i} spectra. The  derived rms values are 0.16 K  for $^{12}$CO per 0.79  km s$^{-1}$ and  0.09 K for both $^{13}$CO  and  C$^{18}$O per 0.83 km s$^{-1}$.

\subsection{\rm{Radio continuum} }
\label{sec:continuum}

To calculate excitation temperature from the HINSA features, background continuum temperature $T\rm _c$ is needed. The Milky Way background continuum temperature is estimated to be $\sim 0.8$ K in the L band (e.g.,\ Winnberg et al.\ 1980). Total $T\rm_c$ containing contribution from the cosmic microwave background (2.7 K; Fixsen 2009) and the Milky Way is estimated to be 3.5 K, but $T\rm _c$ of 3.5 K is only valid for lines of sight toward high Galactic latitudes and  $T\rm _c$ in the Galactic plane is seriously affected by continuum sources, e.g., H II regions. We  adopted 1.4 GHz continuum data from the Continuum H{\sc i} Parkes All-Sky Survey (CHIPASS; Calabretta, Staveley-Smith, \& Barnes 2014) with an angular resolution of 14.4$\arcmin$ and a sensitivity of 40 mK to derive $T\rm_c$. The CHIPASS covers the sky south of declination +25\degree\ that corresponds to $-180\degree<l<68\degree$ in the Galactic plane. In  $68\degree<l<175\degree$, continuum data from CGPS with a rms of $\sim$ 0.3 mJy beam$^{-1}$ at 1420 MHz were utilized.  

\section{Procedures for HINSA identification and Gaussian fitting}
\label{sec:analysis}

\begin{figure*}[t]
  \centering
   \includegraphics[width=0.49\textwidth]{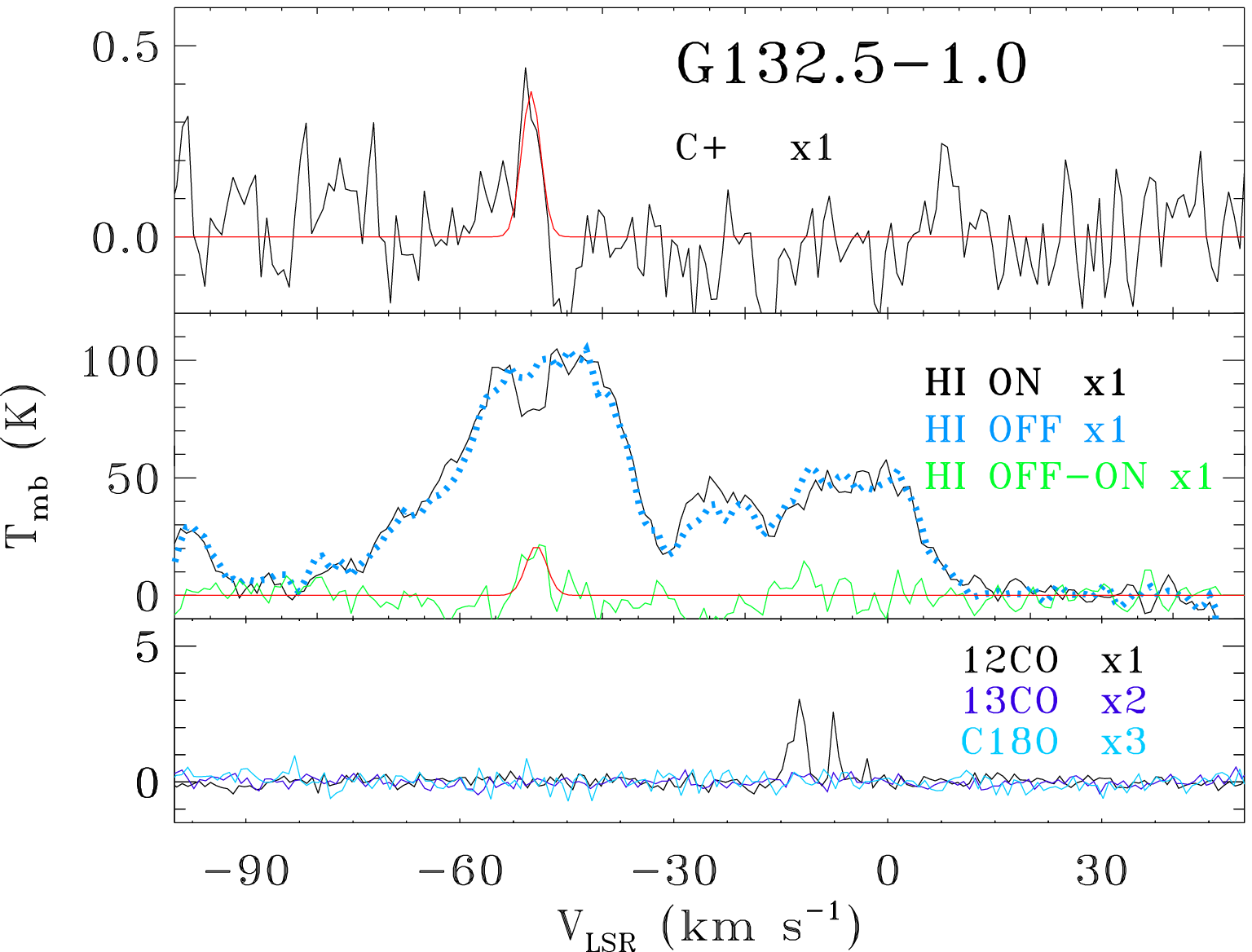}
    \includegraphics[width=0.49\textwidth]{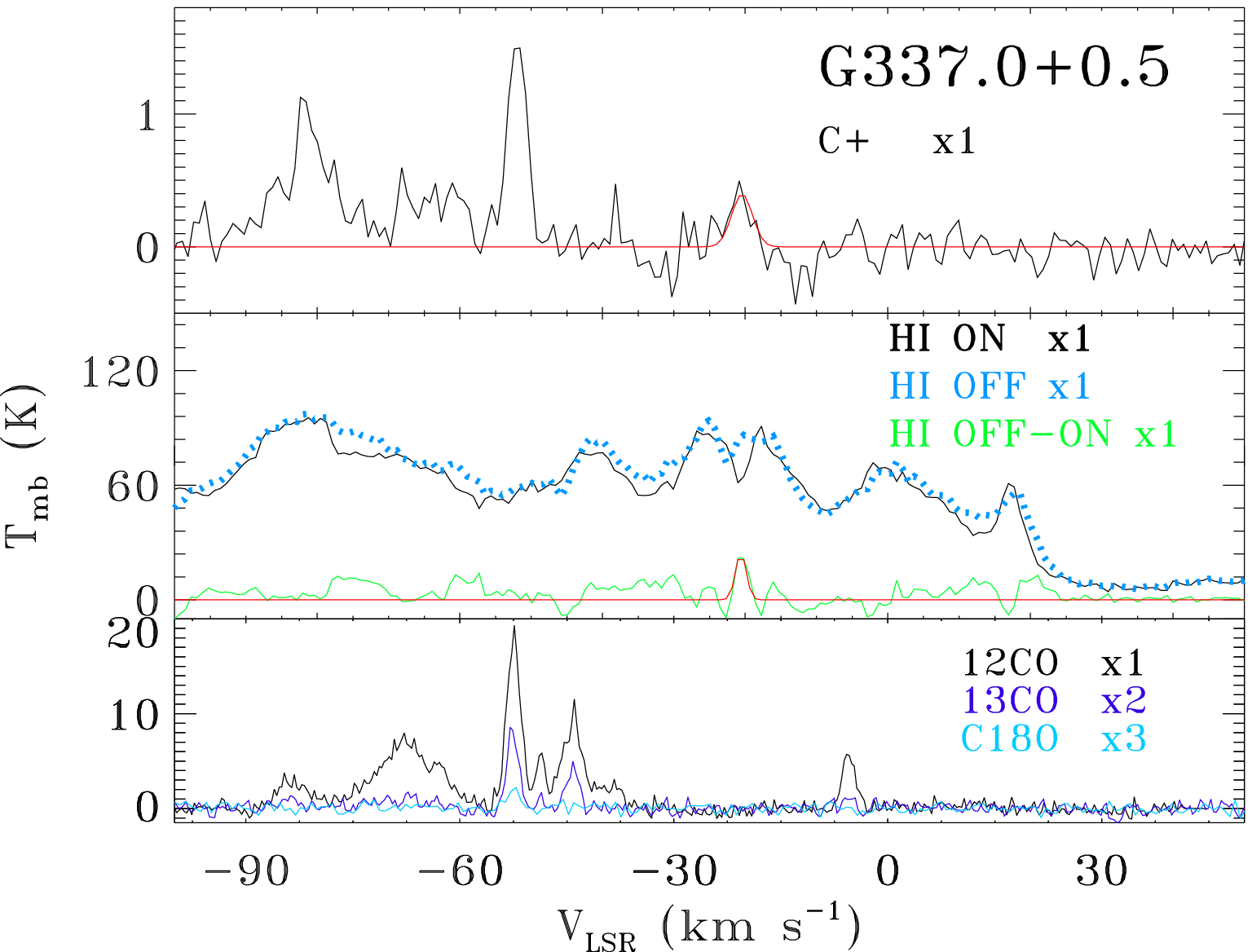}
    \includegraphics[width=0.49\textwidth]{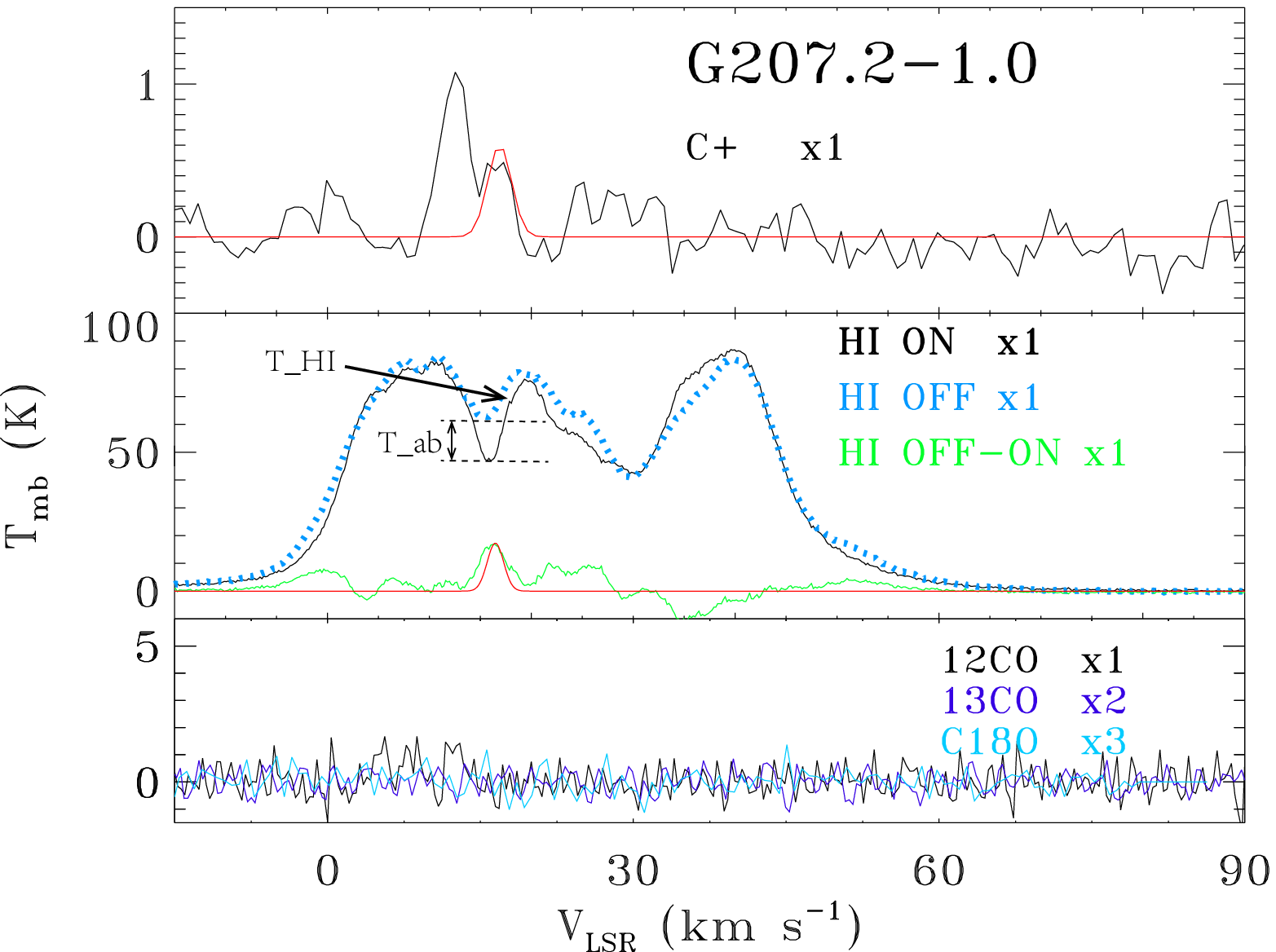}
    \includegraphics[width=0.49\textwidth]{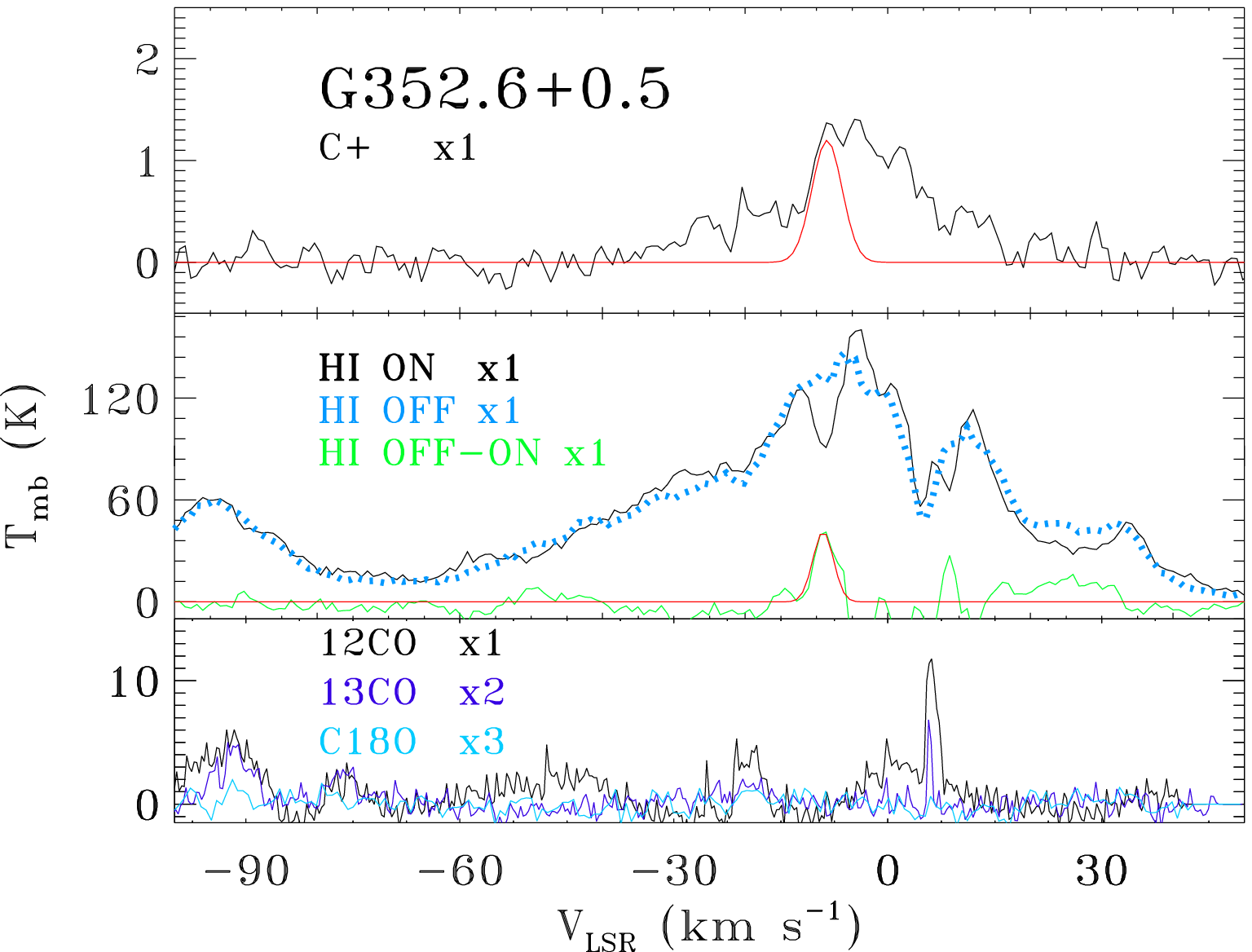} 
\caption{   C$^+$,\hi, and CO spectra toward  G132.5-1.0, G337.0+0.5, G207.2-1.0, and G352.6+0.5. They are shown in top, middle, and bottom panel of each plot, respectively. In the top panel, the red solid line represents the gauss fit of the C$^+$ emission profile of DMG component. In the middle panel, \hi On and OFF profile are derived from ON and around each sightline, respectively.  The green solid line represents the residual profile derived from OFF-ON.  The red solid line represents the gauss fit of the OFF-ON profile around the velocity of the DMG component. In the middle panel toward G207.2-1.0, $T\rm_{\hi}$ and $T\rm_{ab}$ in Equation \ref{eq:excitation} are labeled.  }

\label{Fig:fitting}
\end{figure*}
As shown in Figure  \ref{Fig:fitting}, the  relations between H{\sc i}, C$^+$, and CO are complicated. For example, the cloud at V$\rm_{lsr}$ -52 km s$^{-1}$ for G337.0+0.5 has \hi, $^{12}$CO, and \cp emission. In contrast, the cloud at -44 km s$^{-1}$ for G337.0+0.5 has only \hi and $^{12}$CO, but no \cp emission. Our focus in this study is DMG clouds that have \cp emission  with corresponding HINSA features, but without CO emission. 

\begin{figure*}[t]
  \begin{centering}
  \includegraphics[width=0.49\textwidth]{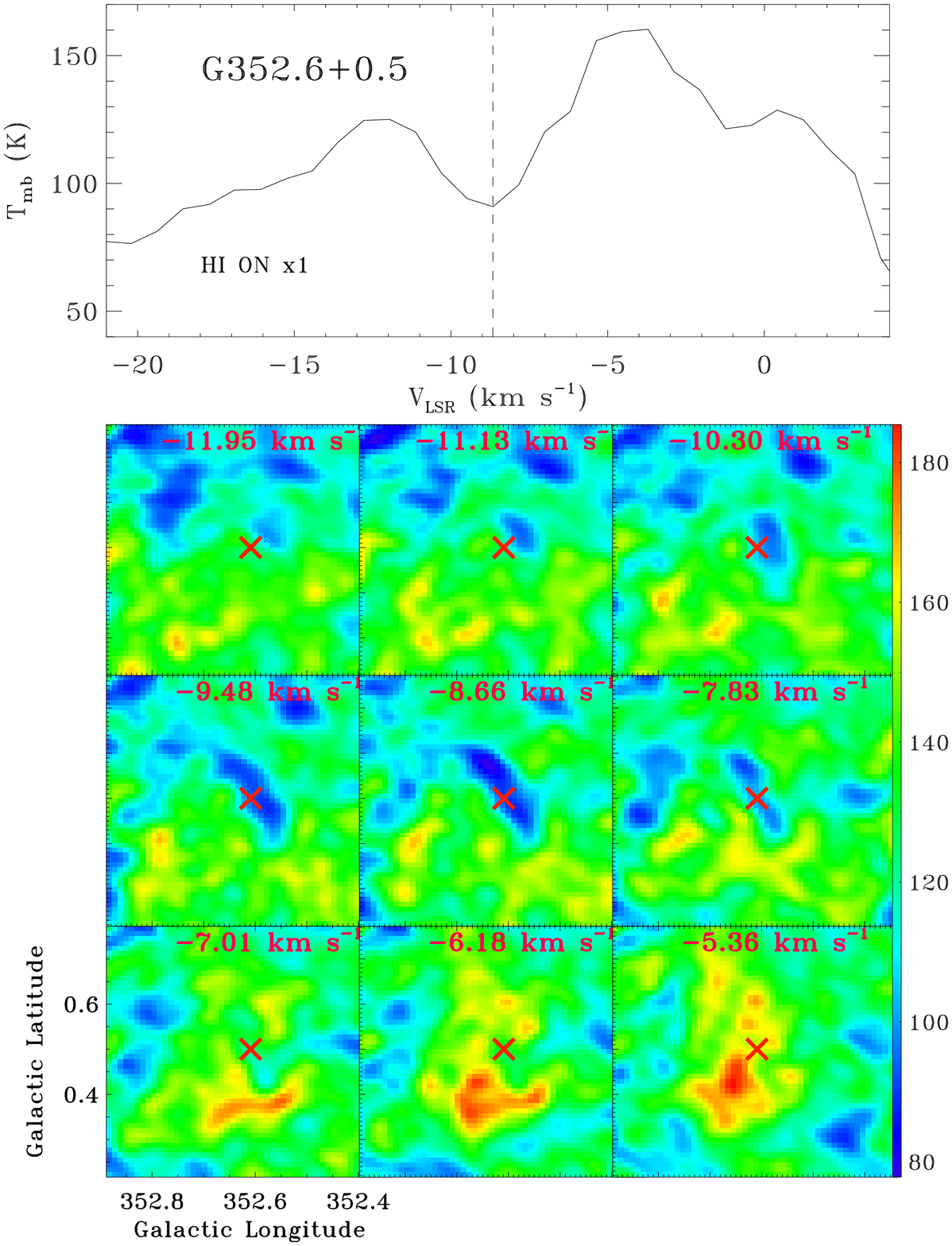}
  \includegraphics[width=0.49\textwidth]{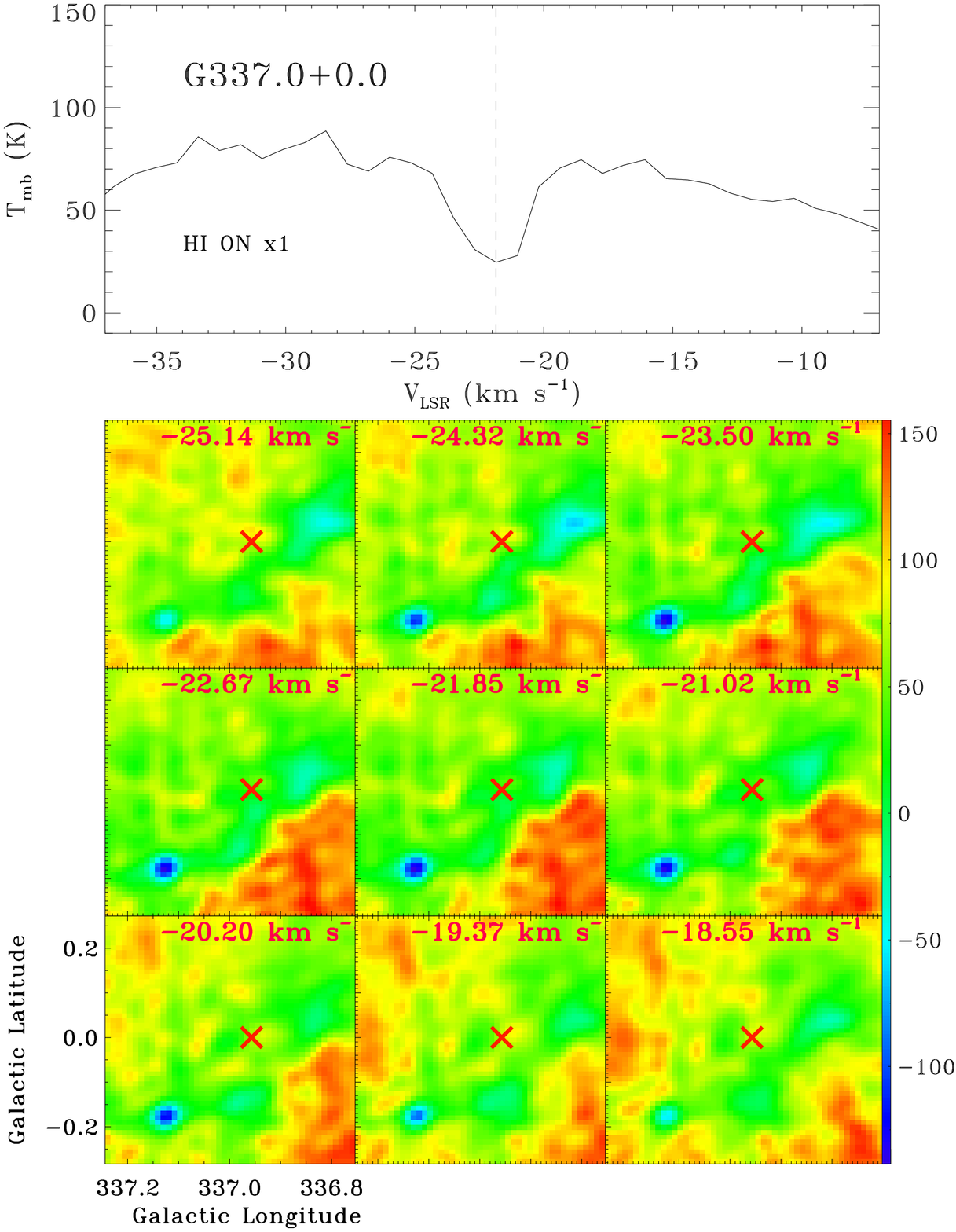}
  \end{centering}
\caption{ \hi spectrum  and \hi channel map around absorption velocity toward G352.6+0.5 and G337.0+0.0, respectively. The dashed line on top of two \hi spectra profiles represents the central velocity of the absorption line. Red crosses below two channel maps represent positions toward G352.6+0.5 or G337.0+0.0.  HINSA with real H{\sc i} absorption of foreground cold \hi cloud around $\sim -8.7$ km s$^{-1}$ is clearly seen toward G352.6+0.5. Non-HINSA with pseudo-absorption caused by relatively high brightness temperatures in neighbor velocity channels around $\sim -21.9$ km s$^{-1}$ is shown toward G337.0+0.0. }
\label{Fig:channelmap}
\end{figure*}

The first step is to identify DMG-HINSA candidates showing \cp emission  and \hi depressions but no obvious CO emission. We found 377 such candidates toward 243 sightlines out of a total of  452  in the GOT C+ program by eye. The candidates were further filtered by the following procedures:
\begin{enumerate}
\item  Depression features are common in Galactic \hi spectra. They can be caused  by  temperature fluctuations, gap effects between multiple emission lines, absorption toward continuum sources, or cooling through collision with \h2 (HINSA) as  described in Section \ref{sec:introduction}. We checked \hi channel map around the depression velocity to ascertain whether a \hi depression feature is HINSA. A HINSA cloud should appear as a colder region than its surroundings in the \hi channel map at its absorption velocity. Moreover, the colder region should be visible in maps of adjacent velocity channels ($\ge 2$). Checking the channel map is necessary because non-HINSA features, with an obvious \hi spectral depression feature, are common. Examples of HINSA and non-HINSA features are shown in Figure  \ref{Fig:channelmap}. We rejected more than half of \hi depression features as fake HINSA features  after this inspection. 

\item  After visual inspection, we employed a quantitative inspection of the absorption to weed out confusion originating from temperature fluctuations. The \hi spectrum toward the GOT C+ sightline was labeled the ON spectrum.  Background \hi  emission arising from behind the foreground absorption cloud was derived through averaging  spectra of nearby positions around the absorption cloud and was labeled the  OFF spectrum. The nearby positions were selected from regions with HI emission contiguous with the ON position and at about 5 arcmin from the cloud boundary.   An absorption signal in the ON spectrum is seen as an emission feature in the OFF-ON spectrum. The component in the residual OFF-ON spectrum is contributed by foreground cold HI cloud. (e.g., around $-50$ km s$^{-1}$ toward G132.5-1.0 of Figure \ref{Fig:fitting}). \hi ON spectra in velocity ranges where Galactic \hi emissions are absent (e.g., $V \ge$ 60 km s$^{-1}$ or $V \le$  -20 km s$^{-1}$ in the \hi spectrum of G207.2-1.0) were chosen to calculate 1$\sigma$ rms. \hi OFF-ON signals with signal-to-noise  (S/N) greater than 3.0 were identified as absorption lines. 

\item  The rms values in different \cp spectra vary owing to different integration time. Spectral ranges without obvious signals were chosen to calculate 1$\sigma$ rms.  The typical 1$\sigma$ rms of \cp is listed in Table \ref{table:rms}. The rms values of different \cp spectra vary by as much as a factor of 1.5. Those \cp signals  with S/N greater than 2.5 were identified as \cp emission lines, considering the generally weaker \cp emission  for clouds without CO emission. 

\end{enumerate}

\begin{table}
\caption{Information of \hi and  \cp lines}
\centering
\begin{tabular}{c c c c}
  \hline
  \hline
Line & $\Delta V$ & Typical $1 \sigma$ rms & Detection threshold  \\
        & km/s      &  K                  & of S/N \\
\hline
\hi   & 0.8  & 1.6-2  & 3.0 \\
\cp  & 0.8  & 0.08 & 2.5 \\
\hline
\end{tabular}
\label{table:rms}
\end{table}
Most spectra are complicated and are far from isolated Gaussian components. Decomposition is necessary. The outcome of the fitting is sensitive to initial inputs, especially the number of Gaussian components and the central velocities of individual components. We developed an IDL code to do Gaussian decomposition. In the code, the number of Gaussian components was automatically determined by the method presented in Lindner et al.\ (2015). The key is the solution of derivatives of the spectra. A regularization method is introduced. It is difficult to define a suitable regularization parameter, which controls the smoothness of the derivations of the spectra. We chose a coarse regularization parameter that may introduce extraneous components. A visual check of all components was performed to remove obviously unreliable components. The estimated parameters of Gaussian components were input as initial conditions into the Gaussian fitting procedure gfit.pro, which was adopted from the Millennium Arecibo 21 cm absorption-line survey (Heiles \& Troland 2003), to give final fitting parameters of decomposed \hi and \cp components.

We first fitted Gaussians to the \hi OFF-ON spectra around HINSA velocity because the HINSA components are easily recognized.  In most cases,  \hi OFF-ON spectra can be fitted with only one Gaussian component. In other cases, two  components were used, and no case of three components was needed. The derived \hi parameters were used as initial conditions for \cp emission fitting.   Examples of Gaussian decomposition for \hi  and \cp spectra are shown in  Figure  \ref{Fig:fitting}. 

The derived components were further filtered based on line widths. We required that an emission line should have at least two channels, corresponding to 1.6 km s$^{-1}$ in \cp and \hi spectra.  

 A final check was necessary  to determine whether the observed \hi gas  can produce the observed \cp emission alone; details are  in Section \ref{sec:cpanalysis}. Finally, we ended up with 36 DMG clouds with relatively clearly visible HINSA features   and Gaussian components. 


\section{Derivation of cloud parameters}
\label{sec:results}

\subsection{\rm{Galactic spatial distribution}}
\label{sec:galacticdistribution}

Kinematic distance was derived based on the Milky Way rotation curve (Brand \& Blitz 1993). The galactocentric radius, $R$, for a cloud with Galactic longitude, $l$, latitude, $b$, and radial velocity along line of sight, $V\rm_{los},$  is given by
\begin{equation}
R=R_{\odot}\frac{V\r{_R}}{V_{\odot}}\frac{V_{\odot}\mathrm{sin}(l)\mathrm{cos}(b)}{V\r{_{los}}+V_{\odot}\mathrm{sin}(l)\mathrm{cos}(b)},
\label{eq:radius}
\end{equation}
where  $V\rm_R$ is orbital velocity at $R$. $V_{\odot}$=220 km s$^{-1}$  is local standard of rest (LSR)  orbital velocity of the Sun at $R_{\odot}$ of 8.5 kpc as recommended by International Astronomical Union (IAU); $V\r{_R}/V_{\odot}=a_1(R/R_{\odot})^{a_2}+a_3$ with $a_1=1.00767, a_2=0.0394$, and $a_3$=0.00712 (Brand \& Blitz 1993).  Then the distance to the cloud, $d$, can be expressed as a function of $R$. In the outer galaxy ( $R  > R_\odot$), the solution is unique, $d=R_\odot \r{cos}(l)/\r{cos}(b) + \sqrt{R^2-R_\odot^2 \r{sin}(l)^2}/\r{cos}(b)$.   In the  inner Galaxy ( $R  < R_\odot$), there exists kinematic distance ambiguity (KDA) with two simultaneous solutions  for a velocity along a line of sight, $d=R_\odot \r{cos}(l)/\r{cos}(b) \pm \sqrt{R^2-R_\odot^2 \r{sin}(l)^2}/\r{cos}(b)$. 

There are three main  resolutions of the KDA: (1) \hi absorption against bright pulsars  (Koribalski et al.\ 1995) or against H II regions with well-known distances (Kolpak et al.\ 2003);  (2)  judgement of different angular extent of the cloud at the near and far kinematic distances (e.g., Clemens et al.\ 1988 ); and (3) the HINSA method.  Clouds in the near distance tend to show HINSA features while clouds in the far distance do not  because of the lack of absorption background (Roman-Duval et al.\ 2009, Wienen et al.\ 2015).  A comparison of the optical image with the $^{13}$CO distribution for GRSMC 45.6+0.3 supports this premise (Jackson et al.\ 2002).  

While solutions 1 and 2 are limited to  sources satisfying specific conditions, solution 3 can be applied to more sources. To test the validity of our distance calculation,  we compared our calculated kinematic distance with maser trigonometric parallax distances  for four sources listed in Table 3 of Roman-Duval  et al.\ (2009). Two kinds of distances are consistent within  $\le$ 5\%. We took the near distance value for our sources located in the inner Galaxy. The  distance thus derived was used to calculate the background  \hi fraction $p$ in Equation \ref{eq:excitation} in Section \ref{sec:HINSA}. 

The above distance estimates have the following caveats: (1) There may exist enough background for a cloud to show HINSA, even at the far distance;  for example, such a background can be provided by spiral density waves  (Gibson et al.\ 2002, 2005a).  (2) The existence of cloud-to-cloud velocity dispersion of about 3 \kms (Clemens 1985) adds uncertainty to the one-to-one mapping of distance to velocity.  Streaming motions of  3 \kms\ will introduce an  uncertainty  of $<\sim$ 220 pc for cloud with $(l,b)$=(45\degree, 0\degree) and LSR velocity of  40 \kms.  

\begin{figure}
  \centering
  \includegraphics[width=0.47\textwidth]{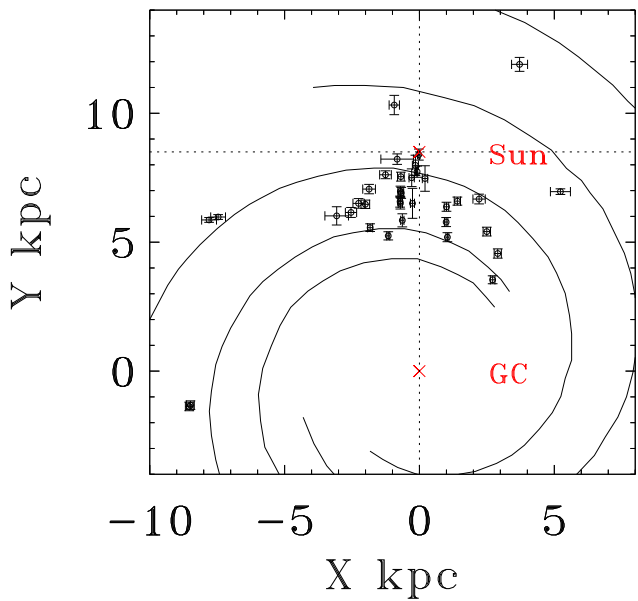}
\caption{Spactial distribution of DMG clouds in the Galactic plane. The positions of the Sun and the Galactic center (GC in the plot) are indicated with red crosses.}
\label{fig:positiondistri}
\end{figure}

Figure \ref{fig:positiondistri} shows the spatial  distribution of 36 DMG clouds in the Galactic plane. Four Galactic  spiral arms  revealed by distributions of star-forming complexes in  Russeil (2003) are also drawn.  It can be seen that most clouds are located between 311\degree\ and 55\degree\ in Galactic longitude. The two ends of the longitude range correspond to tangent directions along Scutum-Crux Arm and Sagittarius Arm, respectively. Selection effect may contribute to this. Foreground clouds preferentially exhibit HINSA features   when they are backlit by warmer \hi emerging  from the Galactic bar and spiral arms.  

\subsection{\rm{Analysis of HINSA}}
\label{sec:HINSA}

The excitation temperature of  cold H{\sc i} absorption cloud can be derived as  (Li \& Goldsmith 2003)
\begin{equation}
T_{\mathrm{ex}}=T_\mathrm{c}+\frac{p\cdot T_{\mathrm{\hi}}-[T_{\mathrm{ab}}/(1-e^{-\tau\rm_{\hi}})]}{1-\tau\rm_f},
\label{eq:excitation}
\end{equation}
where $T\rm_c$ is the background continuum temperature derived from CHIPASS and  CGPS continuum data (Section \ref{sec:continuum}); $p$ is  H{\sc i}  fraction behind the foreground cold cloud;  $T\rm_{\hi}$ is the reconstructed background H{\sc i} brightness temperature without absorption of the foreground cold cloud; and $T\rm_{ab}$ is  the absorption brightness temperature. The temperatures $T\rm_{\hi}$ and $T\rm_{ab}$ are shown in the spectra toward G207.2-1.0 in Figure \ref{Fig:fitting}; $\tau\rm_f$, the foreground H{\sc i} optical depth, was adopted as 0.1; and $\tau\rm_{\hi}$  is the optical depth of \hi in the cold cloud. Infinite $\tau\rm_{\hi}$ results in an upper limit of excitation temperature, $T\rm_{ex}^{upp}$.   Kolpak et al.\ (2002) showed an average optical depth of 1 for clouds in the spatial  range between Galactic radius 4 and 8 kpc. As seen from Figure \ref{fig:positiondistri}, most of our clouds are located in  that  spatial   range.  Thus  it is reasonable to assume $\tau\rm_{\hi}=1$ for  our clouds. The uncertainties of adopting different $\tau\rm_{\hi}$ are discussed further in Section \ref{sec:darkgas}.  

Galactic H{\sc i} spatial  distribution and  positions of DMG clouds are necessary for  calculating $p$. The Galaxy was divided into a set of concentric rings, with a galactocentric radius $R$ and radius width $\Delta R$ = 1 kpc. The H{\sc i} surface density $\Sigma(r)$ of each concentric ring was assumed to be constant and distributed as Figure 10 in Nakanishi \& Sofue (2003). The maximum galactocentric radius of the Galaxy was chosen as 25 kpc. The  spatial information  derived in the Section \ref{sec:galacticdistribution} was applied here. The \hi fraction behind foreground cold cloud $p=\int\r{^{behind}_{cloud}}\Sigma(r)dr/\int\r{^{entire}_{sightline}}\Sigma(r) dr$, where $\int\r{^{entire}_{sightline}}\Sigma(r) dr$ is the total integrated \hi surface density along a sightline  and the $\int\r{^{behind}_{cloud}}\Sigma(r)dr$ is the integrated \hi surface density behind the cloud. 

\begin{figure}
  \centering
  \includegraphics[width=0.49\textwidth]{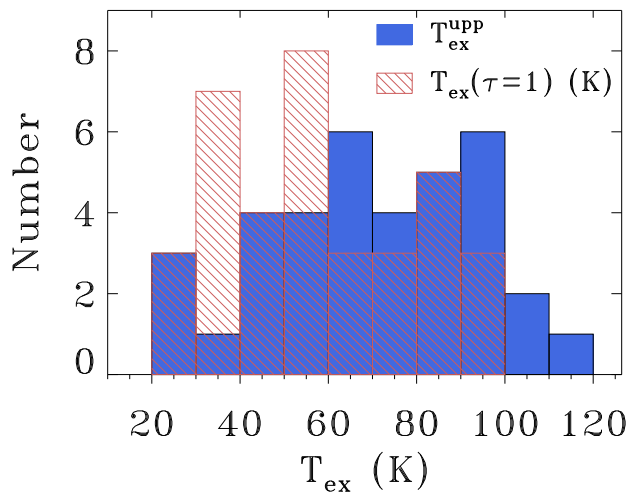}
  \vspace{10pt}
\caption{Histogram of excitation temperature distribution.  Blue rectangles represent the distribution of upper limits to the excitation temperature.  Rectangles with filled red lines represent the distribution of excitation temperature when optical depth $\tau\rm_{\hi}=1$. }
\label{temperaturedistri}
\end{figure}

Derived $T\r{_{ex},(\tau\rm_{\hi}=1)}$ and  $T\rm_{ex}^{upp}$ are shown in column (4) and (5) of Table \ref{tab:table2}, respectively. Excitation temperature distributions of DMG are shown in Figure \ref{temperaturedistri}. $T\r{_{ex},(\tau\rm_{\hi}=1)}$ ranges from 20 to 92 K, with a median value of 55 K.  This median value is comparable to the observed median temperature of 48 K for 143  components of cold neutral medium (Heiles \& Troland (2003)), which were decomposed from emission/absorption spectra  toward 48 continuum sources. Moreover, this median value  is consistent with the calculated temperature range of $\sim 50 - 80$ K  in the CO-dark H$_2$ transition zone in Wolfire et al.\ (2010). The derived lowest $T\r{_{ex},(\tau\rm_{\hi}=1)}$ is 20.3 K for G028.7-1.0. 

The uncertainties of $T\rm_{ex}$  result from $p$ are associated with two aspects. The first is our  adoption of the average \hi surface density in each concentric ring is ideal. This is idealized for two reasons. Firstly, and probably more importantly, is the presence of the localized \hi structure, some of which is associated with the very dark gas we are studying; excess \hi associated with this structure can lie in front of or behind the HINSA. Secondly, such a smooth \hi distribution on large scales is idealized because it neglects such things as spiral structure. The second is the distance ambiguity of the cloud, which may cause a twice the uncertainty. For instance, the near and far distance of G025.2+0.0 is 2.4 kpc and 13.0 kpc. The values of $p$  are 0.86 and 0.59, resulting in $T\rm_{ex}$ of 52.9 K and 28.7 K, respectively.  As we discussed in Section \ref{sec:galacticdistribution}, we prefer the near distance due to \hi absorption feature in our sources. Thus the derived \hi excitation temperature is  an upper limit because of our adoption of the near distance.

With the condition of $h\nu/kT\rm_{ex} \ll 1$, \hi column density $N(\r{\hi})$ is related to \hi optical depth $\tau\rm_{\hi}$ and excitation temperature $T_\r{ex}$  through
\begin{equation}
N(\r{\hi})=1.82\times 10^{18} T_\r{ex} \int \tau\rm_{\hi}d \upsilon  \  \r{cm}^{-2}. 
\label{equ:colhi}
\end{equation}

We derived $N(\r{\hi})$ by adopting $T\r{_{ex},(\tau\rm_{\hi}=1)}$ and $\tau\rm_{\hi}=1$, where  $T_\r{ex}$ is excitation temperature of the cloud. The values of $N(\r{\hi})$ are shown in column (8) of Table \ref{tab:table2}, assuming $\tau\rm_{\hi}=1$. The median value of $N(\r{\hi})$ is 3.1$\times 10^{20}$ \cmtwo. As seen in Equation \ref{eq:excitation}, $T\rm_{ex}$ depends on $\tau\rm_{\hi}$.  The uncertainty in $\tau\rm_{\hi}$ would strongly affect $N(\r{\hi})$ and  the DMG fraction as seen in Section \ref{sec:darkgas}.

The HINSA angular scale, $\Delta \theta$, can be measured from \hi channel maps. Though most HINSA have  a complex nonspherical structure, we used a geometric radius to model the HINSA region in \hi channel map. For a cylinder structure, we chose the width as cloud diameter.  For some HINSA clouds without a clear boundary, there may exist larger uncertainties.  Combining with the calculated distance  $d$ in Section \ref{sec:galacticdistribution}, we can determine the spatial  scale of cloud  $L\rm_{\hi}$=$\Delta \theta\cdot d$. \hi volume density can then be calculated through $n(\r{\hi})=N(\r{\hi})/L\rm_{\hi}$. The derived $n(\r{\hi})$ are shown in column (6) of Table \ref{tab:table2} with a median value of 34 \cmthr, which is consistent with the typical CNM volume density, $n(\rm \hi)_{CNM} \sim 56$ \cmthr (Heiles \& Troland 2003).

\subsection{\rm{Analysis of C$^+$}}
\label{sec:cpanalysis}

\cp is one of the main gaseous forms of carbon elements in the Galactic ISM. It exists in ionized medium, diffuse atomic clouds,  and diffuse/translucent  molecular gas regions where the phase transition between atomic and molecular gas happens (e.g., Pineda et al.\ 2013).  The \cp 158 $\mu$m line intensity, a major cooling line of the CNM, is sensitive to physical conditions. This line is an important tool for tracing star formation activity and ISM properties in the Milky Way and galaxies (e.g., Bosellietal 2002, Stacey et al.\ 2010).  

The \cp 158 $\mu$m line is mainly excited by collisions with electrons, atomic  hydrogen, and  molecular hydrogen.  Collisional rate coefficient (s$^{-1}$/cm$^{-3}$) with electrons  is  $\sim$ 100  times larger than that with atomic and molecular hydrogen because of the advantage of  Coulomb focusing (Goldsmith et al.\ 2012). The \cp emission from ionized gas contributes only 4\% of the total \cp\ 158 $\mu$m flux in the Milky Way (Pineda et al.\ 2013). Our selected clouds have  $T\rm_{ex}$ less than 100 K and should be cold neutral medium (CNM) without high percentage of  ionization.  In  neutral region, \hi and \h2  dominate collisions with \cp.  \cp intensity can be given by (Goldsmith et al.\ 2012)

\begin{align}
\begin{split}
I(\rm C^+) & = 3.279\times 10^{-16} \chi\r{_H(C^+)}\times \\
                &   \bigl[ \frac{N(\r{\hi})}{1+0.5(1+\frac{n\r{_{cr}(\hi)}}{n\r{_{\hi}}})e^{(\Delta E/kT\r{_k})}} + \\
               &      \frac{2N(\r{H_2})}{1+0.5(1+\frac{n\r{_{cr}(H_2)}}{n\r{_{H_2}}})e^{(\Delta E/kT\r{_k})}} \bigr] \ ,
\label{eq:cpinten}
\end{split}
\end{align}
where $\chi\rm_{H}(C^+)$ is \cp abundance  relative to hydrogen; $\chi\r{_{H}(C^+)=5.51\times 10^{-4} exp}(-R\r{_{gal}}/6.2)$, is valid for spatial  range 3 kpc $< R\rm_{gal}<$ 18 kpc (Wolfire et al.\ 2003); and $\chi\rm_{H}(C^+)=1.5\times 10^{-4}$ was adopted outside that range. The parameters $n\mathrm{_{cr}}$(H{\sc i}), $n\rm_{cr}(H_2)$ are  critical densities of H{\sc i} and H$_2$, respectively;   $n\rm_{cr}(\rm{\hi})= 5.75\times 10^4/(16+0.35T^{0.5}+48T^{-1}) $ \cmthr and  $\rm \i{n}_{cr}(H_2)=2\i{n}\rm_{cr}(\hi)$ were adopted from Goldsmith et al.\ (2010);  $n\rm_{\hi}$ and $n\rm_{H_2}$ are volume densities of \hi and \h2, respectively;  $\Delta E/k,$  the transition temperature between  $^2P_{3/2}$-$^2P_{1/2}$ of \cp, is 91.26 K; and $T\rm_k$ is gas kinetic temperature. It is equivalent to $T\rm_{ex}$ of H{\sc i} because H{\sc i}  21 cm emission is always in local thermodynamic equilibrium (LTE)  in gas with density $\gtrsim 10 $ cm$^{-3}$ due to low \hi critical density ($\sim$ $10^{-5}$ cm$^{-3}$) . 

We estimated $\rm  \i{n}(H_2)=\i{N}(H_2)/L_{H_2}$, where $L\r{_{H_2}}$ is the diameter of \h2 layer in cloud and $L\r{_{H_2}}=L\r{_{\hi}}$ was adopted as already discussed in Section \ref{sec:introduction}.   Thus $N$(\h2) and $n$(\h2) can be determined from Equation \ref{eq:cpinten}, and the results  are shown in column (7) and (9) of Table \ref{tab:table2}, respectively. The median value of  $n$(\h2) is  2.3$\times10^2$ \cmthr. The median value of $N$(\h2) is 2.1$\times 10^{21}$ \cmtwo. 

\begin{figure}
  \centering
  \includegraphics[width=0.45\textwidth]{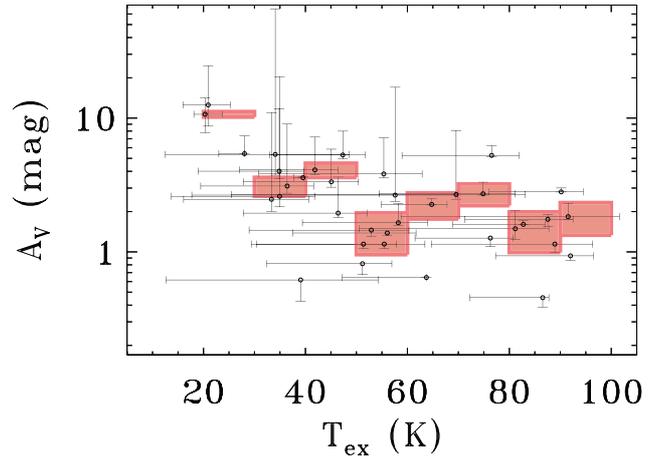}
  \vspace{10pt}
  \caption{ Relation between visual extinction and excitation temperature. Red rectangles indicate median visual extinctions for excitation temperature bin of 10 K. The physical widths and heights of the rectangles are 10 K and 1 mag, respectively. }
\label{Fig:Av-T}
\end{figure}

Visual extinction is connected with total proton column density through,  $A\r{_V}=5.35\times10^{-22}[N(\r{\hi})+2N(\r{H_2})]$ mag, assuming a standard Galactic interstellar extinction curve of $R\r{_V}=A\r{_V}/E(B-V)=3.1$ (Bohlin, Savage \& Drake 1978). The corresponding visual extinction values toward each source are shown in column (11) of  Table \ref{tab:table2}. In Figure \ref{Fig:Av-T}, we plot $A\rm_V$ as a function of $T\rm_{ex}$. It is clear that $A\rm_V$ has a decreasing trend when $T\rm_{ex}$ increases.

The ratio between $L\r{_{H_2}}$ and $L\r{_{\hi}}$ is a key relation during the above calculation, but may vary for clouds with different visual extinction and different PDR models. We took another value $L\r{_{H_2}}/L\r{_{\hi}}=0.8$, which is the possible lower value of PDR with $A\rm_V<0.2 mag$, to estimate the uncertainty. With ratios of 1.0 and 0.8, the maximum differences of N(H$_2$), A$\rm_V$, and DMG fraction (Section \ref{sec:darkgas}) are 10\%, 10\%, and 5\%, respectively. Thus the value of the ratio $L\r{_{H_2}}/L\r{_{\hi}}$ does not affect the physical parameters associated with H$_2$ too much.  

\section{ Inferred DMG cloud properties}
\label{sec:DMGcloud}

\subsection{\rm{Observed properties of dark gas clouds}}
\label{sec:darkgas}

Physical properties and spatial  distribution of DMG are fundamental quantities that  affect our understanding of the transition between diffuse atomic clouds and dense molecular clouds.  If extra \h2 not traced by CO is needed  to explain the observed \cp intensity, the cloud is considered a DMG cloud. 

Following Equation (7) in L14, the mass fraction of DMG in the cloud is defined as,
\begin{equation}
f\r{_{DMG}}=\frac{2N(\mathrm{CO\text{--} dark ~H_2})}{N(\mathrm{\hi})+2N\mathrm{(H_2)}},
\label{equ:fraction}
\end{equation}
where $N\mathrm{(H_2)}=N(\mathrm{CO\text{--}dark ~H_2})+N(\mathrm{CO\text{--}traced ~H_2})$. In this paper,  $N(\rm CO\text{--}traced\ H_2)$ is set to 0  due to absence of CO detection for our samples. 

\begin{figure*}
  \centering
  \includegraphics[width=0.45\textwidth]{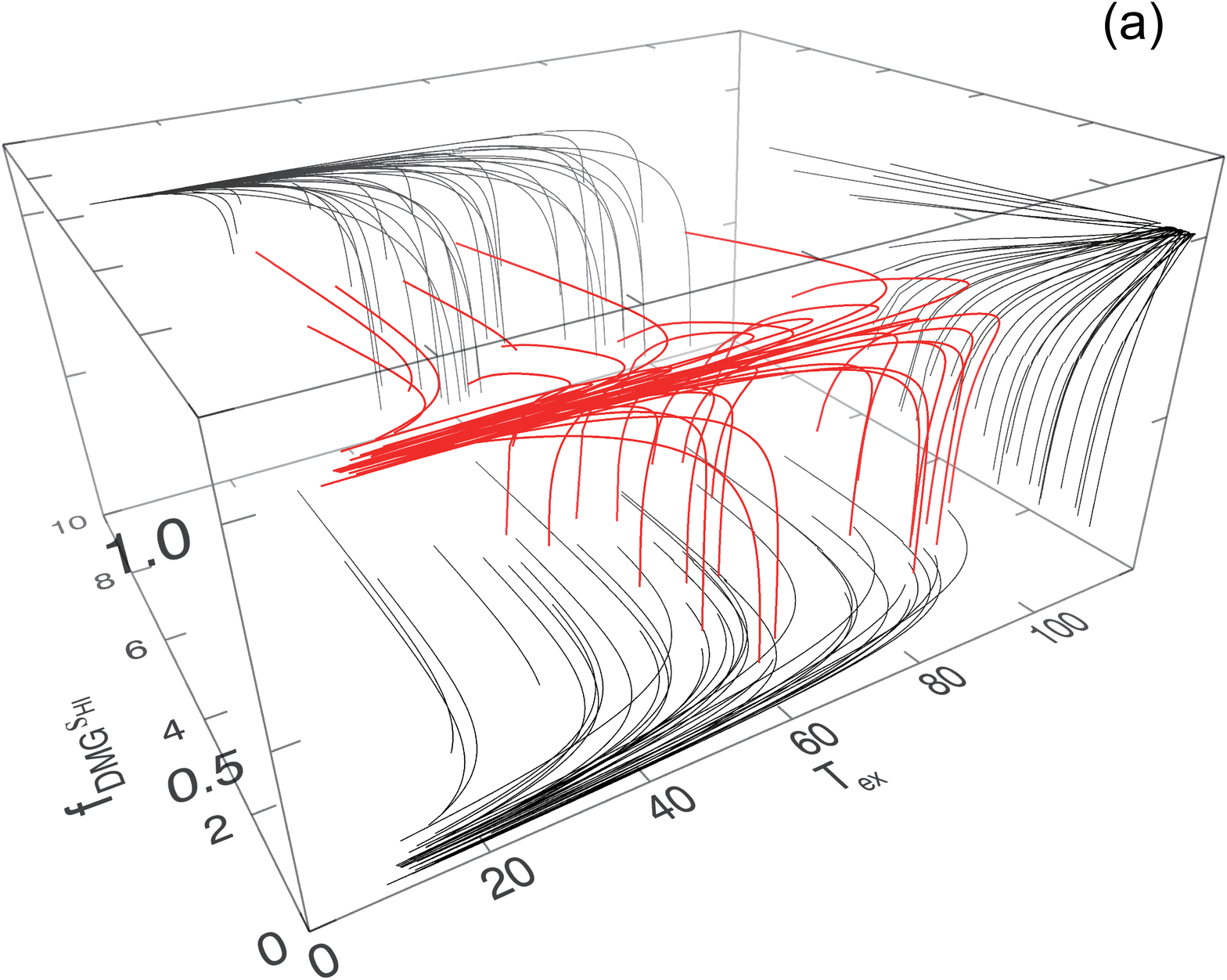}
   \includegraphics[width=0.45\textwidth]{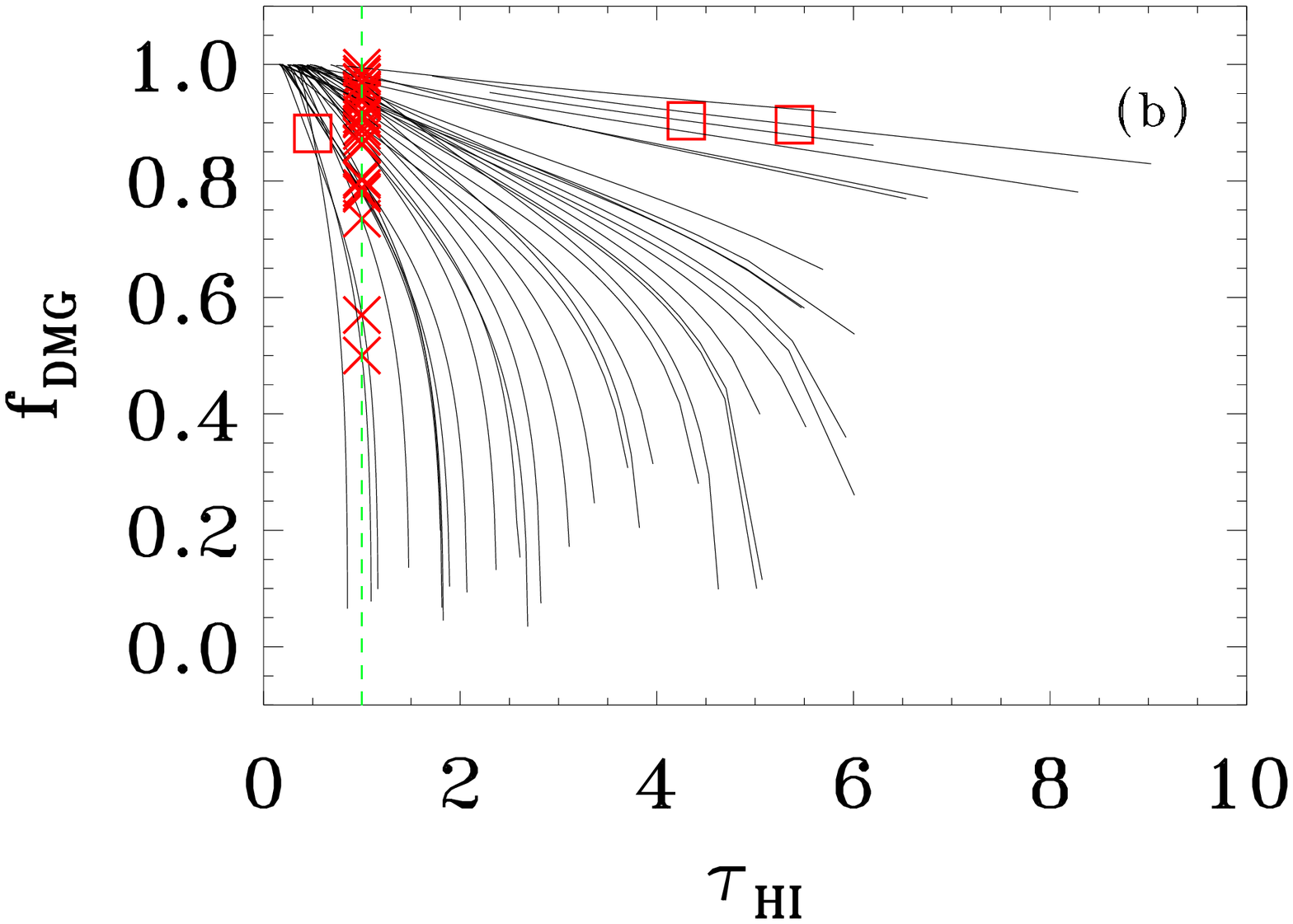}
    \includegraphics[width=0.45\textwidth]{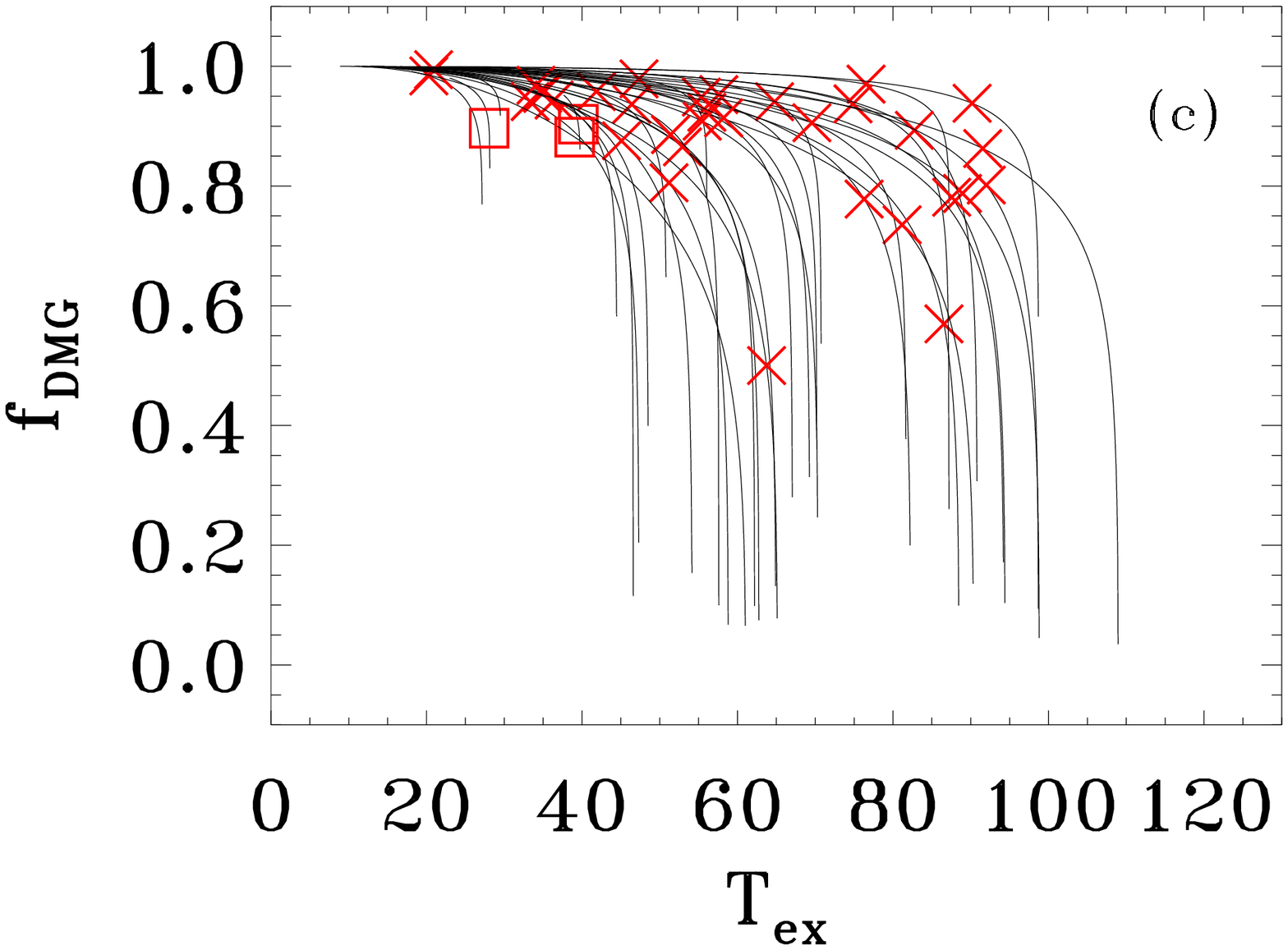}
     \includegraphics[width=0.45\textwidth]{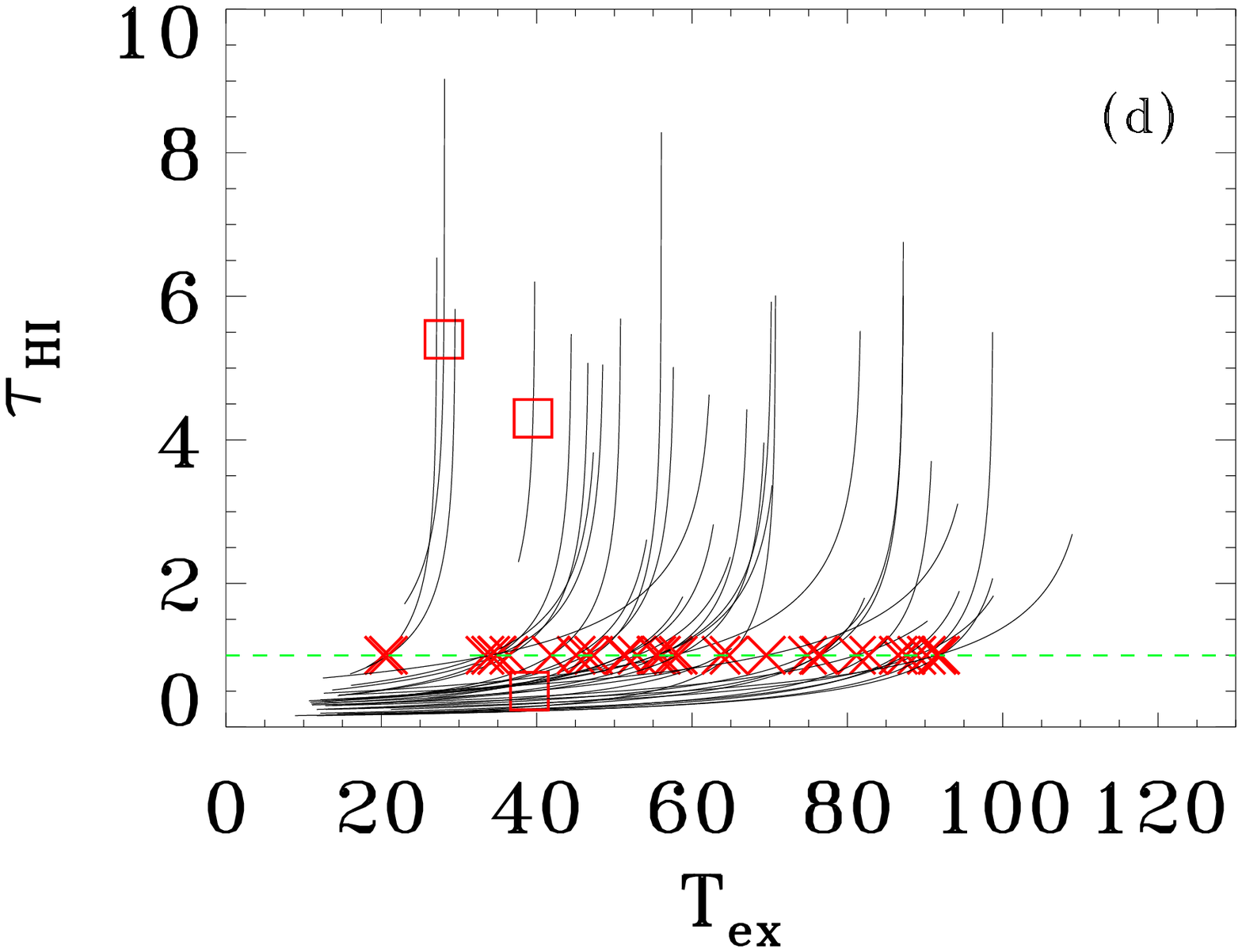}
  \caption{Relation curves between $f\rm_{DMG}$, $\tau\rm_{\hi}$, and $T\rm_{ex}$. Figure (a)  shows  3D curves with red lines. Relations between each two parameters are projected with black lines.   Figure (b)  shows relation between $f\rm_{DMG}$ and $\tau\rm_{H{\sc i}}$. The green dashed line represents $\tau\rm_{\hi}=1$. The value with $\tau\rm_{\hi}=1$ for  33 sources are indicated with red crosses,  but for G132.5-1.0,  G207.2-1.0, and G347.4+1.0, $\tau\rm_{\hi}=5.4, 4.3$, and 0.5, respectively. They are shownd with red squares. Figure (c)  shows the relation between $f\rm_{DMG}$ and $T\rm_{ex}$. Figure (d)  shows the relation between $\tau\rm_{\hi}=1$ and $T\rm_{ex}$. The meaning of red crosses and blue dashed lines in Figure (c) and (d) are same as in Figure (b).}
\label{fig:uncertainty}
\end{figure*}

The uncertainty of $f\rm_{DMG}$ comes from two aspects. First, measurement and fitting of the \hi and \cp spectra. They were estimated to be less than $\sim 10$\% for all the sources. The second is the uncertainty of adopting $\tau\rm_{\hi}$ of \hi.  As seen in Section \ref{sec:HINSA}, $\tau\rm_{\hi}$ of \hi greatly affects the \hi column density, and thus $f\rm_{DMG}$. It is necessary to investigate available parameter space.  The parameters are constrained by the following three conditions:
\vspace{-4pt}
\begin{enumerate}
\item  $T\rm_{ex} > 0 $ K.  
\item  $N\rm_{H_2} \ge 0$ cm$^{-2}$.
\item  The derived extinction $A\rm_V$ $\le$ $A\rm_V(dust)$, where $A\rm_V(dust)$ is the total Galactic dust extinction along the sightline. We adopted extinction values from all sky dust extinction database (Schlafly \& Finkbeiner 2011), in which dust extinction was derived through analyzing colors of stars E(B-V) of Sloan Digital Sky Survey with a reddening ration $A\rm_V$/E(B-V)=3.1.  
\end{enumerate} 
\vspace{-4pt}
The relations between $f\rm_{DMG}$, $T\rm_{ex}$, and $\tau\rm_{\hi}$ for 36 sources are shown in Figure \ref{fig:uncertainty}. It is worthwhile to note that the upper values of $\tau\rm_{\hi}$ are overestimated and lower values of $\tau\rm_{\hi}$ are underestimated as $A\rm_V(dust)$ is the total value  along the sightline of each source. The parameter $A\rm_V(dust)$ contains contributions from CO-traced molecular gas at other velocities besides those with dark gas.  

According to Equation (\ref{eq:excitation}),  higher $T\rm_{ex}$ is required to produce a fixed absorption strength when $\tau\rm_{\hi}$ increases. This is reflected in Figure \ref{fig:uncertainty}(d). According to Equation (\ref{equ:colhi}), a bigger $\tau\rm_{\hi}$ produces larger $N(\rm \hi)$. 

In Figure \ref{fig:uncertainty}(b), we present $f\rm_{DMG}$ versus $\tau\rm_{\hi}$. When $\tau\rm_{\hi}$ increases, $f\rm_{DMG}$ decreases to a nonzero minimum value.  This can be understood as follows. \cp is mainly excited by collisions with \hi and \h2. According to Equation (\ref{eq:cpinten}), for a fixed C+ intensity, increasing $N(\rm \hi)$ and $T\rm_k$ ($T\rm_k$=$T\rm_{ex}$ was adopted) increases the contribution of \hi collision to \cp emission, requiring decreasing contribution from \h2 collision, thus decreasing the \h2 column density and decreasing the \h2 fraction in the cloud. The lower limits of $\tau\rm_{\hi}$ for all sources are less than 1.0 except for G132.5-1.0 and G207.2-1.0, which have a $\tau\rm_{\hi}$ range of (1.7,9.0) and (2.3,6.2), respectively.   For these two sources,  we apply median value $\tau\rm_{\hi}=5.4$ and 4.3, respectively.   As seen in Figure \ref{fig:uncertainty}(c) and \ref{fig:uncertainty}(d), this  selection does not affect $T\rm_{ex}$ and $f\rm_{DMG}$ too much as they are in narrow value ranges, (23.0, 28.2) K and (0.83, 0.98) for G132.5-1.0, (37.7, 39.8) K and (0.86, 0.95) for  G207.2-1.0; $\tau\rm_{\hi}=0.5$ is applied for G347.4+1.0 due to an upper limit of 0.85.  For other  33 sources, we apply $\tau\rm_{\hi}=1.0$. Although this selection  is arbitrary, we argue that it is reasonable for two reasons. The first reason is that  averaged \hi optical depth between the Galactic radius 4 and 8 kpc is around 1.0 (Kolpak et al.\ 2002). The second reason is that the changes from 0.5 to 1.5 of $\tau\rm_{\hi}$ strongly affect $f\rm_{DMG}$ value for only  three sources. For other sources, the values of $f\rm_{DMG}$ have a minimum  of $\ge 0.6$ in this $\tau\rm_{\hi}$ range, implying a weak dependence of $\tau\rm_{\hi}$ in the range of [0.5,1.5].  Thus we take a $\tau\rm_{\hi}$ range of [0.5,1.5] to represent the total uncertainty since uncertainties of $\tau\rm_{\hi}$ are much greater than measurement and fitting uncertainties.

\begin{figure}
  \centering
  \includegraphics[width=0.45\textwidth]{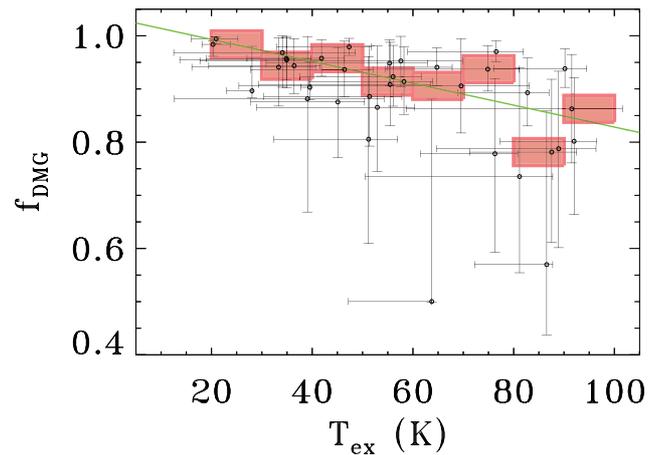}

\caption{Relation between DMG fraction  and excitation temperature.  Median values of $f\rm_{DMG}$ per 10 K are shown with red rectangles. The width and height of the rectangles are 10 K and 0.05, respectively. Green solid curve represents best linear fitting for median values. }
\label{Fig:frac-spin}
\end{figure}

The relation between $f\rm_{DMG}$ and  $T\r{_{ex},(\tau\rm_{\hi}=1)}$ is shown in Figure  \ref{Fig:frac-spin}.   The relation between DMG fraction and gas excitation temperature can be described well by an empirical relation,

\begin{equation}
f\r{_{DMG}}=-2.1\times 10^{-3}T\r{_{ex},(\tau\rm_{\hi}=1)}+1.0 . 
\label{equ:fracandtemp}
\end{equation}

The decreasing trend of $f\rm_{DMG}$ toward increasing $T\rm_{ex,(\tau\rm_{\hi}=1)}$ is clear. This result  is consistent with that in  Figure 7 of Rachford et al.\ (2009). With the FUSE telescope, Rachford et al.\ (2009) derived the total molecular hydrogen $N$(H$_2$) and rotational temperature $T\rm_{01}$ directly through UV absorption of \h2 toward bright stars. These authors found that molecular fraction $\rm \i{f}=2\i{N}(H_2)/(2\i{N}(H_2)+\i{N}(\hi))$ decreases from $\sim 0.8$ at $T\rm_{01}=45$ K to $\sim 0.0$ at $T\rm_{01}=110$ K with relatively large scatter. Though the decreasing trend between our result and that in Rachford et al.\ (2009) is similar,  $f\rm_{DMG}$ is as high as $7.7\times 10^{-1}$ at $T\rm_{ex}=110$ K in Equation \ref{equ:fracandtemp}, implying a  flatter slope compared to that in Rachford et al.\ (2009). Our results are more  physical meaningful because $N(\r{\hi})$ and $T_{01}$ in Rachford et al.\ (2009) are averaged values along a line of slight.

\begin{figure}
  \centering
  \includegraphics[width=0.45\textwidth]{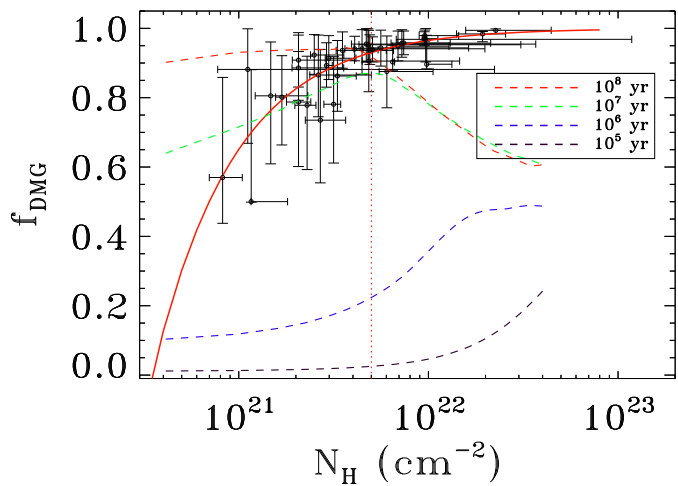}
  \vspace{10pt}
  \caption{ Fraction of DMG vs. hydrogen column density N$_H$.  Results for 36 sources are shown in open circles.  Red solid line shows the best fitting. Dashed lines with different colors represent chemical evolutionary model results from Lee et al.\ (1996).  Vertical dotted red line represents $N\rm_H= 5\times 10^{21}$ cm$^{-2}$. }
\label{fig:f-NH}
\end{figure}

Relation between $f\r{_{DMG}}$ and  $N\rm_H$ is shown in  Figure \ref{fig:f-NH}. It reflects DMG fractions along different extinctions and is investigated in most theoretical papers. The data are fitted with an empirical relation
\vspace{-2pt}
\begin{equation}
f\r{_{DMG}}=1-\frac{3.7\times 10^{20}}{N\rm_H/cm^{-2}}.
\label{equ:fracandav}
\end{equation}

We compared this result with cloud evolutionary model from Lee et al.\ (1996), who incorporated time-dependent chemistry and shielding of CO and H$_2$ in photodissociation clouds.  Lee et al.\ (1996) split the cloud into 43 slabs. We adopted Lee's model through the following procedures. First, we calculated total hydrogen column density $N\rm_H$ and total \h2 column density $N$(H$_2$) at 43 slabs. Then  CO-traced H$_2$ was calculated through $N$(CO-traced H$_2$)= $N$(CO)/$Z$(CO), where $Z$(CO) is CO abundance relative to molecular hydrogen. We derived the DMG column density through $N$(DMG)=$N$(H$_2$)-$N$(CO-traced H$_2$). Finally,  the DMG fraction $f\rm_{DMG}$=2$N$(DMG)/$N\rm_H$.  As already shown in the models of Lee et al.\ (1996),  $Z$(CO) varies significantly under different environments as shown in the chemical models (e.g., Lee et al.\ 1996) and in observations toward diffuse gas clouds (Liszt \& Pety 2012).  We adopted a constant $Z$(CO)= 3.2$\times 10^{-4}$ (Sofia et al.\ 2004) that is an upper limit in the ISM during the calculation. This leads to an upper DMG fraction from the model.

We adopted model 1 in Lee et al.\ (1996), in which all hydrogen were originally in atomic phase. The DMG fractions as a function of hydrogen column density in the age of 10$^5$, 10$^6$, 10$^7$, and 10$^8$ yr are shown in  Figure \ref{fig:f-NH} with dashed lines. It can be seen that our results are consistent with model results at age of $10^7$ yr when $N\rm_H\lesssim 5\times 10^{21}$ cm$^{-2}$ ($A\rm_V \lesssim$ 2.7 mag). When $N\rm_H > 5\times 10^{21}$ cm$^{-2}$, $f\rm_{DMG}$ decreases according to the modeled results of chemical evolution but still increases in our results. This difference persists even we consider data uncertainties. The 36 clouds  were thus divided into two groups: a low extinction group with $N\rm_H\lesssim 5\times 10^{21}$ cm$^{-2}$ and high extinction group with $N\rm_H > 5\times 10^{21}$ cm$^{-2}$. 

Planck Collaboration (2011) found an apparent excess of dust optical depth $\tau\rm_{dust}$ compared to the simulated   $\tau\rm_{dust}^{mod}$ between $A\rm_V$ range of [0.37, 2.5] mag. The $A\rm_V$ value of $\sim 0.37$ mag and  $\sim 2.5$ mag correspond to threshold extinction of H$_2$ self-shielding and threshold extinction of dust shielding for CO, respectively. When $A\rm_V >   2 .5$ mag, the CO abundance increases, resulting in a deceasing DMG fraction as expected from the chemical evolutionary model predictions at the age of $10^7$ yr  in Figure \ref{fig:f-NH}. If the CO luminosity is too weak to be observed, this would lead to an increasing curve  when $A\rm_V >  2 .5$ mag. Actually, Liszt \& Pety (2012) found patchy CO emission in DMG regions with higher CO sensitivity. 

In order to estimate  CO abundance limits in high extinction group ($A\rm_V >2.7$ mag) clouds, we assumed optically thin and LTE of CO. These two assumptions are reasonable owing to no detection of  CO and $\rm \i{T}_{ex}$/5.56 K $\gg$ 1. $^{12}$CO column densities were derived through $N(^{12}\r{CO})=4.8\times 10^{14}\int T\r{_b} d\upsilon~\rm cm^{-2}$. We used a rms of $T\rm_b=0.6$ K and  velocity  resolution of 0.35 km s$^{-1}$ in our CO spectra. An upper limit of $^{12}$CO column density $N$(CO)=$1.0\times 10^{14}$ cm$^{-2}$ implies an upper CO abundance relative to \h2 $Z\rm_{CO}^{upp}=\i{N}(CO)/\i{N}(H_2)= 2.1\times 10^{-6}$ for $A\rm_V> 2.7$ mag; $Z\rm_{CO}^{upp}$ is  $6.6\times 10^2$ times smaller than the canonical value of $3.2\times 10^{-4}$ in the Milky Way (Sofia et al.\ 2004).

Our assumption of optically thin emission of CO  in low $A\rm_V$ clouds is mostly empirical. This assumption can be quantified as the following. We smoothed the data to RMS of 0.44 K per 0.7 km s$^{-1}$ . For a cloud with modest opacity at $T\rm_{ex} =10$ K, $T\rm_{bg}=2.7$ K, $\tau(\rm CO)$ = 1, the derived antenna temperature (50\% main beam efficiency) is 2.1 K, which is way above our RMS threshold.

Thus we conclude that, clouds in high extinction group ($A\rm_V >2.7$ mag) are  CO poor molecular clouds. The formation of these clouds are discussed in Section \ref{sec:cloud_assembly}.

\subsection{\rm{Comparison between clouds in low and high extinction groups}}
\label{sec:low-high-compare}

\begin{figure}
  \centering
  \includegraphics[width=0.47\textwidth]{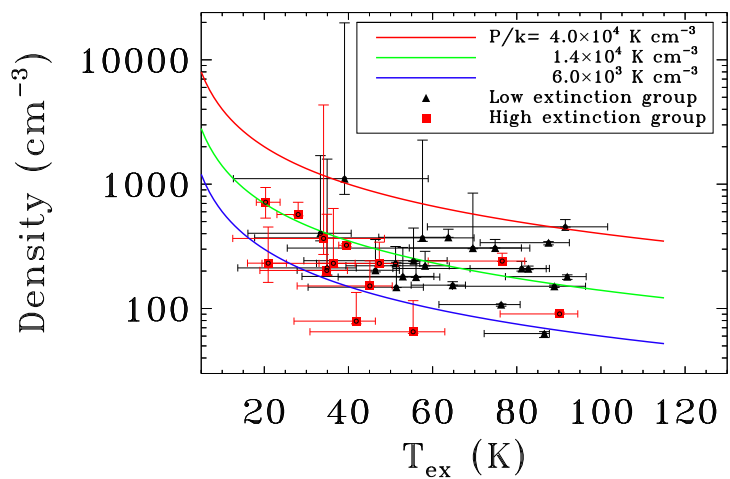}
  \vspace{10pt}
\caption{ Relation between volume density and excitation temperature. Blue, green, and red lines represent pressure  $P/k$ of 6000, 1.4$\times 10^4$, 4 $\times 10^4$  K \cmthr, respectively.  }
\label{Fig:n-T}
\end{figure}

We  plotted the total gas volume density $\rm \i{n}{_{gas}}=\i{n}\rm{_{\hi}}+\i{n}\rm{_{H_2}}$ as a function of $T\rm_{ex}$ for 36 sources in Figure \ref{Fig:n-T}. Typical thermal pressure  $P\rm_{th}$ of $6\times 10^3$ K \cmthr in Galactic radius of 5 kpc (Wolfire et al.\ 2003), $P\rm_{th}$ of $1.4\times 10^4$ K \cmthr  near the Galactic center (Wolfire et al.\ 2003), and auxiliary  $P\rm_{th}$ of $4\times 10^4$ K \cmthr are also shown.   Median densities for  the low extinction ($A\rm_V \le 2.7$ mag) and high extinction ($A\rm_V >2.7$ mag) groups are 212.1 and 231.5 cm$^{-3}$, respectively.  The median excitation temperatures for the low extinction ($A\rm_V \le 2.7$ mag) and high extinction ($A\rm_V >2.7$ mag) groups are 64.8 and 41.9 K, respectively.  Densities in these two groups are comparable but excitation temperatures are relatively lower in high extinction ($A\rm_V >2.7$ mag) group, resulting lower thermal pressures in this group.  We discuss the implication for cloud formation in Section \ref{sec:cloud_assembly}.

\section{Discussion}
\label{sec:discussion}
\subsection{\rm{Assembly of molecular clouds}}
\label{sec:cloud_assembly}

Molecular clouds can be formed either  directly from neutral medium or by assembling pre-existing, cold molecular clumps. The first scenario is commonly accepted (e.g., Hollenbach \& McKee 1979). The second scenario is outlined by Pringle et al.\ (2001), who proposed that clouds formed out of pre-existing, CO-dark, molecular gas. Compared to the first scenario, the second scenario allows for fast cloud formation in a  few Myr, which was suggested by observations (Beichman et al. 1986; Lee, Myers \& Tafalla 1999; Hartmann et al. 2001). The key problem for the second evolutionary scenario is how molecular gas can exist before it is collected together.  Pringle et al.\ (2001) argued that the pre-existing molecular gas should be  cold ($< 10$ K) and was shielded from photodissociation by patch dust with $A\rm_V$ of $\sim 0.5$ mag  (White et al.\ 1996; Berlind et al.\ 1997; Gonzalez et al.\ 1998; Keel \& White 2001). Under $A\rm_V$ of $\sim 0.5$ mag, H$_2$ should exist substantially while CO  is in very low abundance.  This is because the self-shielding threshold of  $A\rm_V$=0.02 and 0.5 mag (Wolfire et al. 2010) are widely considered as a condition for maintaining a stable population of abundant H$_2$ and CO gas, respectively. Listz et al. (2012) detected strong CO(1-0) emission (4-5 K) in regions with equivalent visual extinction less than 0.5 mag. Two obvious possibilities are 1) the CO gas is transient. Such gas may also have been seen in Goldsmith et al. 2008, who detected a ~40\% of the total CO mass in Taurus in regions with low to intermediate Av (mask 0 and 1 in their terminology).  2) Such CO gas lies in a highly clumpy medium with lower apparent averaged extinction when photons travel through lower density interclump medium. When  small agglomerations of molecular gas are compressed and heated by shock, e.g., in a spiral arm, they become detectable. This scenario is supported by observations of GMC formation in spiral arms (Dobbs et al.\ 2008) and simulations of molecular clouds formation (e.g., Clark et al.\ 2012). 
 
 In Section \ref{sec:darkgas}, we showed that clouds in high extinction group ($A\rm_V >2.7$ mag) are not consistent with chemical evolutionary model of the first scenario. The upper limit of CO abundance in this group is $6.6\times10^2$ times smaller than the typical value in the Milky Way. We suggest the CO-poor feature can be explained if clouds in the high extinction group ($A\rm_V >2.7$ mag) are formed through coagulation of pre-existing molecular, CO-poor clumps. The clouds should be in the early stage of formation.  According to chemical evolutionary model, CO can reach an abundance of $2\times 10^{-5}$ in $10^5$ yr at $A\rm_v=2.7$ mag if all hydrogen is locked in H$_2$ before the cloud formation (Lee et al.\ 1996). Thus the cloud age may be constrained to be less than $1.0\times 10^5$ yr after cloud assembly.  
 
 Moreover, the obvious differences seen in linewidth-scale relation, excitation temperature distribution, and non-thermal/thermal ratio relations for clouds in the low extinction group ($A\rm_V \le 2.7$ mag) and high extinction group ($A\rm_V >2.7$ mag) in Section \ref{sec:low-high-compare}, are possible pieces of evidence to support the cloud formation under the second scenario.

\subsection{\rm{\hi contributes  little in explaining dark gas}}
Dark gas is the gas component that is not detected with either \hi or CO emission but is clearly seen from the  excess of $A\rm_V$ compared to $N\rm_H$ (Reach et al.\ 1994). We focus on DMG, but as  pointed out in Planck Collaboration (2011), $N$(H{\sc i}) could be underestimated with the optically thin approximation and an excitation temperature that is too high. Atomic \hi may contribute as much as $50\%$ mass for the excess according to their estimate. Fukui et al.\ (2015) investigated the \hi optical depth $\tau\rm_{\hi}$ and reanalyzed all sky Planck/IRAS dust data in high galactic latitudes ($|b|>15$\degree). They derived  2-2.5 times higher \hi densities than that with optical thin assumption. They implied that optically thick cold \hi gas may dominate dark gas in the Milky Way.

In this paper, we introduced HINSA as an effective tool to constrain $\tau\rm_{\hi}$. Though $\tau\rm_{\hi}=1$ is applied for  33 clouds, it does not affect the  conclusion much that DMG dominated the cloud mass for $0.5\le \tau\rm_{\hi}\le 1.5$. 

Another objection against H$_2$ dominating dark gas in Fukui et al.\ (2015) was that, crossing the timescale of $\le 1$ Myr of local clouds is an order of magnitude smaller than H$_2$ formation timescale $2.6\times 10^9/n\rm_{\hi}$ yr (Hollenbach \& Natta 1995; Goldsmith \& Li 2005) for typical clouds ($n\sim 100$ \cmthr). This is not a problem if we adopt the assumption in Section \ref{sec:cloud_assembly} that molecular clouds are formed by assembling pre-existing molecular gas.  

 Within our sample of clouds with C+ emission and HI self-absorption, the molecular gas seems to be the dominant component regardless of their individual excitation temperatures, optical depth, and their lack of CO emission.  Our conclusion is in line with the direct observational result in the Perseus by Lee et al.\ (2015).

\subsection{\rm{Galactic impact}}

Grenier et al.\ (2005) indicated that dark gas mass is comparable to that of CO-traced molecular gas in the Milky Way. Our results suggest that H$_2$ dominates the dark gas.  In a previous study, L14  obtained H{\sc i} intensity by integrating over a velocity range centered around their $V\rm_{LSR}$ defined by the \cp (or $^{13}$CO) line. Moreover, they adopted an optically thin assumption and a constant kinematic temperature of 70 K. To compare with L14, we applied these treatments to DMG clouds in this study. Results from these treatments differ from our results by an average factor of $0.55^{+4.61}_{-0.83}$ for total visual extinction $A\rm_V$ and differ by an average factor of $0.04^{+0.89}_{-0.21}$ for DMG fraction. The symbol ``$+$'' means maximum value of underestimate and ``$-$'' means maximum value of overestimate.  The actual DMG content detected with previous treatments and method here may differ a little. 

The detections in this study are limited for DMG for two reasons. First, DMG clouds without HINSA feature  are common in the Milky Way. Second,  \cp emission  in some DMG clouds may be hard to identify. 

We estimated quantitatively the detection limits in this study. To be detected under the sensitivities of this study,  the excitation temperature should be lower than the background emission temperature. The detection requirement of  \cp brightness temperature is 0.25 K (2.5 $\sigma$) .   As seen in Equation \ref{eq:cpinten}, \cp intensity strongly depends on kinetic temperature $T\rm_k$.  To produce a \cp intensity of $3.2\times10^{-1}$ K km s$^{-1}$ ($T\rm_b^{peak}=0. 3$ K and FWHM of 1.0 km s$^{-1}$),  it  requires a $N$(\h2)=$9.0\times 10^{19}$ \cmtwo under $T\rm_k=70.0$ K and  $N$(\h2)=$2.2\times 10^{21}$ \cmtwo under $T\rm_k=20.0$ K, assuming $n\rm_{H_2}=1.0\times 10^3$ \cmthr. Thus a large fraction of cold, diffuse DMG clouds in the Milky Way may be undetectable as C$^+$ emission is under the conditions specified in this paper.  

\section{Summary}
\label{sec:summary}
In this paper, we have carried out a study of the DMG  properties in the Galactic plane by combining physical properties derived from \cp survey of $Hershel$, international $\rm\hi$ surveys, and CO surveys. The HINSA method was used to determine  \hi excitation temperature, which is assumed to be constant  in previous works (e.g., Langer et al.\ 2014). Our conclusions include 

\begin{enumerate}
\item Most DMG clouds are distributed between the Sagittarius arm and Centaurus arm in the Milky Way.  We argue that this is caused by sample selection with HINSA features, which can be produced only when background temperature is stronger than excitation temperature of foreground cloud. 

\item \hi excitation temperatures of DMG clouds vary in a range between 20 and 92 K with a median value of 55 K, which is lower than assumed 70 K in Langer et al.\ (2014). Gas densities vary from $6.2\times10^1$ to $1.2\times 10^3$ \cmthr with a median value of $2.3\times 10^2$ \cmthr. 

\item  DMG dominates dark gas in a wide range of \hi optical depth $\tau\rm_{\hi}$ and excitation temperature $T\rm_{ex}$. 

\item   The  \hi optical depth $\tau\rm_{\hi}$ can exist in a wide parameter range without significantly affecting the global relations between DMG fraction,  \hi column density, and \hi excitation temperature.

\item  Under the constraint of $\rm^{12}CO$ sensitivity of 0.44 K per 0.7 km s$^{-1}$ in this paper, the  relation between $f\rm_{DMG}$ and excitation temperature can be described by a linear function, $f\rm_{DMG}=-2.1\times 10^{-3}\i{T}_{ex}+1.0$, assuming \hi optical depth of 1.0. 

\item  The relation between $f\rm_{DMG}$ and total hydrogen column density $N\rm_H$ can be described by $f\rm_{DMG}=1-3.7\times 10^{20}/\i{N}\rm_H$.  When $N\rm_H$ $\le 5.0\times10^{21}$ cm$^{-2}$, this curve is consistent with the time-dependent chemical evolutionary model at the age of $\sim 10$ Myr.  The consistency between the data and chemical evolutionary model breaks down when $N\rm_H> 5.0\times 10^{21}$ \cmtwo.

\item We discovered  a group of clouds with high extinction ($A\rm_V > 2.7$ mag), in which an upper CO abundance of $2.1\times 10^{-6}$ relative to \h2 is two orders magnitude smaller than canonical value in the Milky Way. This population of clouds cannot be explained by the chemical evolutionary model. They may be formed through the agglomeration of pre-existing molecular gas in the Milky Way. 
\end{enumerate}

It is worthwhile to note that the definition of DMG strongly depends on the sensitivity of CO data. In this paper, this value is 0.44 K per 0.7 km s$^{-1}$ for $\rm^{12}CO$ emission. More sensitive data of CO as well as other molecular tracers, e.g., OH, toward these clouds are necessary to constrain CO abundance further and to investigate physical properties of molecular gas in these clouds. 


\section*{Acknowledgments}

This work is supported by Technology under State Key Development Program for Basic Research (973 program) No.\  2012CB821802 and National Key Basic Research Program of China ( 973 Program )  2015CB857100, the China Ministry of Science, and the Guizhou Scientific Collaboration Program (\#20130421). We are grateful to the anonymous referee for his/her constructive suggestions, which have greatly improved this paper. The authors would  like to thank Pei Zuo for helping Delingha CO observations, John Dickey for discussing \hi self-absorption, Xiaohu Li for discussing chemical evolutionary model.  We thank the Pineda team for providing C$^+$ and CO data. Part of CO data were observed with the Delingha 13.7 m telescope of the Qinghai Station of Purple Mountain Observatory. We appreciate the help of all the staff members of the observatory during the observations.   

\bibliographystyle{aa} 
\bibliography{references}

\clearpage

\begin{table*}
\caption{Derived parameters.}
\centering
\begin{tabular}{l l l l l l l l l l l  }
  \hline
  \hline
    (1)    &   (2)          &  (3)     & (4)                         &  (5)                       &  (6)       &  (7)                &  (8)                 &  (9)   & (10)  & (11)  \\
ID         &  Source     &   $\Delta V$  &  $T\rm_{ex,(\tau\rm_{\hi}=1)}$  &  $T\rm_{ex}^{upp} $ & $n(\rm\hi)$ & $n$(\h2)        & $N\rm_{\hi}$ & $N\rm_{H_2}$ &  $f\rm_{DMG}$ & $A\rm_V$     \\
           &                   &           &                              &                    & $10^2$ & $10^2$  & $10^{20}$         &  $10^{21}$            &                            &      \\
           &                   &      km s$^{-1}$     &   K                           &     K                     & \cmthr  & \cmthr  & \cmtwo         &  \cmtwo                          &           &   mag      \\
\hline
   &   &   &    &   &      &   &   &    &   &           \\    
1 &  G011.3+0.0 &       2.03 &      57.6 &      81.8 &       0.33 &      3.4 &        2.3 &        2.3 &     0.95 &     2.7  \\ 
2 &  G017.4+1.0 &       4.57 &      34.9 &      44.5 &       0.17 &      1.9 &        3.1 &        3.5 &     0.96 &     4.0  \\ 
3 &  G020.0+0.0 &       2.11 &      86.6 &      95.2 &       0.38 &       0.25 &        3.5 &        0.2 &     0.57 &     0.45  \\ 
4 &  G025.2+0.0 &       3.44 &      52.9 &      67.5 &       0.43 &      1.4 &        3.5 &        1.1 &     0.87 &     1.5  \\ 
5 &  G028.7-1.0 &       7.90 &      20.3 &      27.2 &       0.23 &      6.9 &        3.1 &        9.5 &     0.98 &    11  \\ 
6 &  G036.4-0.5 &       2.48 &      46.4 &      57.7 &       0.24 &      1.8 &        2.2 &        1.6 &     0.94 &     2.0  \\ 
7 &  G036.4-1.0 &       1.92 &      64.8 &      70.8 &       0.17 &      1.4 &        2.4 &        1.9 &     0.94 &     2.3  \\ 
8 &  G038.9-1.0 &       4.47 &      36.4 &      46.7 &       0.25 &      2.1 &        3.2 &        2.6 &     0.94 &     3.1  \\ 
9 &  G050.4+1.0 &       3.15 &      34.9 &      48.7 &       0.18 &      1.9 &        2.1 &        2.2 &     0.95 &     2.6  \\ 
10 &  G073.6+1.0 &       4.07 &      33.3 &      47.9 &       0.45 &      3.6 &        2.6 &        2.1 &     0.94 &     2.5  \\ 
11 &  G132.5-1.0 &       3.45 &      28.1 &      28.2 &      1.1 &      4.6 &       10 &        4.4 &     0.90 &     5.4 \\ 
12 &  G207.2-1.0 &       1.90 &      39.5 &      39.8 &       0.57 &      2.7 &        6.3 &        2.9 &     0.90 &     3.6  \\ 
13 &  G288.7-0.5\_1 &       2.18 &      47.3 &      56.1 &       0.095 &      2.2 &        2.0 &        4.7 &     0.98 &     5.3  \\ 
14 &  G288.7-0.5\_2 &       3.85 &      41.9 &      50.9 &       0.064 &       0.7 &        3.1 &        3.5 &     0.96 &     4.1  \\ 
15 &  G288.7+0.0 &       1.77 &      56.1 &      67.3 &       0.26 &      1.5 &        1.9 &        1.2 &     0.92 &     1.4  \\ 
16 &  G305.1-0.5 &       2.54 &      88.9 &     104 &       0.53 &       0.98 &        4.4 &        0.8 &     0.79 &     1.1  \\ 
17 &  G307.7+0.0 &       1.91 &      76.5 &      87.2 &       0.14 &      2.3 &        2.8 &        4.6 &     0.97 &     5.3  \\ 
18 &  G308.9+0.5 &       2.12 &      74.9 &      87.3 &       0.36 &      2.7 &        3.1 &        2.3 &     0.94 &     2.7  \\ 
19 &  G311.5+0.5 &       4.52 &      81.2 &      99.8 &       0.88 &      1.2 &        7.1 &        1.0 &     0.74 &     1.5 \\ 
20 &  G312.8-1.0 &       1.79 &      90.2 &      98.8 &       0.11 &       0.8 &        3.1 &        2.4 &     0.94 &     2.8  \\ 
21 &  G315.3+0.0 &       1.93 &      82.8 &      91.2 &       0.40 &      1.7 &        3.1 &        1.3 &     0.89 &     1.6  \\ 
22 &  G319.1+0.0 &       3.31 &      55.4 &      70.3 &        0.064 &       0.59 &        3.6 &        3.3 &     0.95 &     3.8  \\ 
23 &  G319.1+0.5 &       3.17 &      20.9 &      29.5 &        0.026 &      2.3 &        1.3 &       11 &     0.99 &    13  \\ 
24 &  G324.3+0.5 &       2.89 &      51.2 &      62.6 &       0.75 &      1.6 &        2.9 &        0.6 &     0.81 &     0.82  \\ 
25 &  G328.1-1.0 &       8.59 &      45.1 &      55.6 &       0.34 &      1.2 &        7.5 &        2.6 &     0.88 &     3.3  \\ 
26 &  G335.2+0.0 &       4.63 &      34.1 &      62.7 &       0.23 &      3.4 &        3.1 &        4.7 &     0.97 &     5.4  \\ 
27 &  G337.0+0.5 &       1.76 &      55.5 &      71.3 &       0.41 &      2.0 &        1.9 &        0.9 &     0.91 &     1.1  \\ 
28 &  G339.6-0.5 &       2.36 &      51.4 &      64.1 &       0.30 &      1.2 &        2.4 &        0.9 &     0.89 &     1.1  \\ 
29 &  G340.4+0.5\_1 &       3.44 &      76.3 &      85.3 &       0.39 &       0.69 &        5.1 &        0.9 &     0.78 &     1.3  \\ 
30 &  G340.4+0.5\_2 &       4.07 &      87.6 &      97.4 &      1.2 &      2.2 &        6.9 &        1.2 &     0.78 &     1.8 \\ 
31 &  G343.9+0.5 &       1.88 &      92.0 &     101 &       0.60 &      1.2 &        3.3 &        0.7 &     0.80 &     0.93 \\ 
32 &  G343.9+1.0 &       4.70 &      63.7 &      73.8 &      2.5 &      1.3 &        5.8 &        0.3 &     0.50 &     0.64  \\ 
33 &  G346.5-0.5 &       2.28 &      58.2 &      69.6 &       0.35 &      1.9 &        2.6 &        1.4 &     0.91 &     1.7  \\ 
34 &  G347.4+1.0 &       3.49 &      39.1 &      61.0 &      2.3 &      8.7 &        1.3 &        0.49 &     0.88 &     0.59  \\ 
35 &  G352.6+0.5 &       3.39 &      69.6 &      96.4 &       0.53 &      2.5 &        4.6 &        2.2 &     0.91 &     2.7  \\ 
36 &  G353.5+0.0 &       2.56 &      91.5 &     111 &      1.1 &      3.4 &        4.5 &        1.4 &     0.86 &     1.8  \\ 
\hline
\end{tabular}
\tablefoot{ Parameter for DMG. Column (1) is ID number. Column (2) is source name. Column (3) is  full width at half maximum of \hi.  Column (4) is excitation  temperature with assumption of optical depth of 1. Column (5) is upper limit of excitation temperature assuming infinite optical depth. Column (6) is \hi volume density. Column (7) is \h2 volume density. Column (8) is \hi column density. Column (9) is \h2 column density. Column (10) is DMG fraction relative to total hydrogen. Column (11) is total visual extinction. }
\label{tab:table2}
\end{table*}

\end{document}